\documentclass[twocolumn]{aastex63}
\usepackage[utf8]{inputenc}
\usepackage{amsmath}
\usepackage{amssymb}
\usepackage{natbib}
\usepackage{xspace}
\usepackage{graphicx}
\usepackage{rotating} 
\usepackage{subfigure}
\usepackage{textcomp}
\usepackage{float}
\usepackage{hyperref}
\usepackage{savesym}
\savesymbol{tablenum}
\usepackage{siunitx}
\restoresymbol{SIX}{tablenum}

\newcommand{\e}[1]{\ensuremath{^{#1}}\xspace}

\newcommand{\nh}{$N_{\text{H}}$\xspace}

\newcommand{\NHgal}{$N_{\text{H, Gal}}$\xspace}
\newcommand{\ecut}{$E_{\text{cut}}$\xspace}
\newcommand{\aox}{$\alpha_{\text{ox}}$\xspace}

\newcommand{\ka}{K$\alpha$\xspace}

\newcommand{\chidof}{$\chi^{2}/$dof\xspace}

\newcommand{\nustar}{\textsl{NuSTAR}\xspace}
\newcommand{\bat}{\textsl{Swift}/BAT\xspace}
\newcommand{\xrt}{\textsl{Swift}/XRT\xspace}
\newcommand{\xmm}{\textsl{XMM-Newton}\xspace}

\shortauthors{Kamraj et al.}

\begin{document}

\title{The Broadband X-ray Spectrum of the X-ray Obscured Type 1 AGN 2MASX J193013.80$+$341049.5}

\correspondingauthor{Nikita Kamraj}
\email{Contact: nkamraj@caltech.edu}

\author{Nikita Kamraj}
\affiliation{Cahill Center for Astronomy and Astrophysics, California Institute of Technology, Pasadena, CA 91125, USA}

\author{Mislav Balokovi\'c}
\affiliation{Center for Astrophysics $\vert$ Harvard \& Smithsonian, 60 Garden Street, Cambridge, MA 02138, USA}
\affiliation{Black Hole Initiative at Harvard University, 20 Garden Street, Cambridge, MA 02138, USA}

\author{Murray Brightman}
\affiliation{Cahill Center for Astronomy and Astrophysics, California Institute of Technology, Pasadena, CA 91125, USA}

\author{Daniel Stern}
\affiliation{Jet Propulsion Laboratory, California Institute of Technology, Pasadena, CA 91109, USA}

\author{Fiona A.~Harrison}
\affiliation{Cahill Center for Astronomy and Astrophysics, California Institute of Technology, Pasadena, CA 91125, USA}

\author{Roberto J. Assef}
\affiliation{N\'{u}cleo de Astronom\'{i}a de la Facultad de Ingenier\'{i}a y Ciencias, Universidad Diego Portales, Av. Ej\'{e}rcito Libertador 441, Santiago, Chile}

\author{Michael J. Koss}
\affiliation{Eureka Scientific Inc., 2452 Delmer St. Suite 100, Oakland, CA 94602, USA}

\author{Kyuseok Oh}
\affiliation{Department of Astronomy, Kyoto University, Oiwake-cho, Sakyo-ku, Kyoto 606-8502, Japan}
\affiliation{JSPS fellow}

\author{Dominic J. Walton}
\affiliation{Institute of Astronomy, University of Cambridge, Madingley Road, Cambridge CB3 0HA, UK}

\begin{abstract}

\noindent We present results from modeling the broadband X-ray spectrum of the Type 1 AGN 2MASX J193013.80+341049.5 using \nustar, \emph{Swift} and archival \xmm observations. We find this source to be highly X-ray obscured, with column densities exceeding 10$^{23}$ cm$^{-2}$ across all epochs of X-ray observations, spanning an 8 year period. However, the source exhibits prominent broad optical emission lines, consistent with an unobscured Type 1 AGN classification. We fit the X-ray spectra with both phenomenological reflection models and physically-motivated torus models to model the X-ray absorption. We examine the spectral energy distribution of this source and investigate some possible scenarios to explain the mismatch between X-ray and optical classifications. We compare the ratio of reddening to X-ray absorbing column density ($E_{B-V}$/\nh) and find that 2MASX J193013.80+341049.5 likely has a much lower dust-to-gas ratio relative to the Galactic ISM, suggesting that the Broad Line Region (BLR) itself could provide the source of extra X-ray obscuration, being composed of low-ionization, dust-free gas. 

\end{abstract}


\section{Introduction}

Under the unified model of Active Galactic Nuclei (AGN) \citep{antonucci,urry-1995}, differences between Type 1 and Type 2 AGN arise solely from our line of sight viewing angle relative to a toroidal obscuring structure surrounding the central supermassive black hole (SMBH). The dusty molecular torus believed to be responsible for the obscuration seen in Type 2 AGN likely has a clumpy distribution, as evidenced from infrared interferometric observations \citep{tristram-2007,ir-interferometry-2016} and short timescale variability of the line-of-sight column density in some AGN \citep[e.g.,][]{marinucci-2016}. The unified model, while simplistic, has been successful in explaining observations of radio jets, polar ionization cones, and polarization position angle \citep[e.g.,][]{evans-1991,storchi-1992}. 

Optical classification of AGN is based on widths of observed emission
lines. Type 1 AGN contain both broad (FWHM $\gtrsim$ 1000 km s$^{-1}$) and narrow emission lines whereas Type 2 AGN contain only narrow emission lines. In the unified picture, this is due to obscuration of the Broad Line Region (BLR), which is close to the central SMBH, by the molecular torus. Hence in Type 2 AGN, only emission from the Narrow Line Region (NLR) is observed, which is extended on kilo-parsec scales. In Type 1 AGN, the torus is viewed at angles such that the line of sight to the BLR is unobscured.

X-ray classification of AGN is typically based on measurements of the hydrogen column density (\nh). Neutral gas along our line of sight absorbs X-ray continuum photons, with the level of absorption dependent on the column density of gas. The approximate dividing line between unobscured, Type 1 AGN and obscured, Type 2 AGN is at \nh $= 10^{22}$ cm$^{-2}$ \citep{ricci-2017a}. Assuming the simple picture of a static obscuring torus surrounding the AGN is correct, then there should be agreement between optical and X-ray classifications. 

The \bat survey comprises all sources detected by the all-sky 14--195 keV Burst Alert Telescope (BAT) instrument onboard the \emph{Neil Gehrels Swift Observatory} \citep{swift-mission,bat-105-month}, and provides a hard X-ray-selected sample of local AGN that is relatively unbiased to obscuration. The \bat sample therefore offers a unique opportunity to test the unified model of AGN. Within the \bat sample, there is excellent agreement between the optical and X-ray classifications for $\sim$ 95\% of sources \citep{koss-2017}. However, there are a number of unusual sources in the \bat sample that show disagreement between their X-ray and optical classifications. Some narrow-line Type 2 AGN have been found to be X-ray unabsorbed and there is some debate about whether they represent a class of AGN lacking a BLR \citep{panessa-2002,tran-2011,merloni-2014}. In contrast, a number of Type 1 AGN (Sy1 - 1.9) are highly X-ray absorbed \citep[e.g.,][]{shimizu-2017}.

In this paper, we study the spectral properties of an unusual source from the \bat survey, 2MASX J193013.80+341049.5, which is optically classified as a Type 1 AGN, yet is found to be highly X-ray absorbed, with a column density \nh $> 10^{23}$ cm$^{-2}$. The source is a nearby ($z = 0.063$) Seyfert galaxy with a bolometric luminosity $L_{Bol} = 1.3\times10^{45}$ erg s$^{-1}$ \citep{koss-2017}, a $R$-band apparent magnitude of 15.8, and a compact optical morphology. We present results from modeling the broadband X-ray spectrum of 2MASX J193013.80+341049.5 using \nustar, \xmm and \bat observations, and examine the multi-wavelength properties of this source. This paper is structured as follows: in section~\ref{sec:data} we describe the X-ray observations and data reduction procedures; in section~\ref{sec:modeling} we present results from broadband X-ray modeling; section~\ref{sec:multiwavelength} presents analysis of multi-wavelength data on this source, including optical spectra and broadband Spectral Energy Distributions (SEDs); section~\ref{sec:summary} summarizes our results and presents our conclusions.

\begin{table*}[t]
	\centering	
	\caption{Details of the \nustar and \xmm observations of 2MASX J193013.80+341049.5 considered in this work.}
	\renewcommand{\arraystretch}{1.0}
	\label{table:table1}
	\begin{tabular*}{\textwidth}{@{\extracolsep{\fill} }l c c c c c}
		
		\noalign{\smallskip} \hline \hline \noalign{\smallskip}
		
		Mission  & Observation ID & Observation Date & Exposure Time & Count rate \\
		& & & (ks) & (counts s$^{-1}$) \\ \hline
		\\
		\nustar & 60160713002 & 2016-07-19 & 23.4 & 0.142 \\
		\nustar & 60376001002 & 2017-10-10 & 55.4 & 0.120 \\ 
		\\
		\xmm & 0602840101 & 2009-05-16 & 16.9 & 0.288 \\
		
		\\ \hline	
		
	\end{tabular*}
\tablecomments{Observed source count rates are in the 0.5--10 keV band for the \xmm EPIC-pn detector and the 3--79 keV band for \nustar (FPMA).}	
\vspace{10pt}
\end{table*}

\section{X-ray Observations and Data Reduction}\label{sec:data}

\subsection{NuSTAR}

2MASX J193013.80+341049.5 was observed with \nustar in July 2016 for $\sim$ 20 ks as part of the Extragalactic Legacy Surveys program\footnote{https://www.nustar.caltech.edu/page/legacy\_surveys}. A second, deeper \nustar observation was performed in October 2017 for $\sim$ 50 ks. Details of both these observations are presented in Table~\ref{table:table1}. 

We reduced the raw event data from both \nustar modules, FPMA and FPMB \citep{nustar-harrison} using the \nustar Data Analysis Software (NuSTARDAS, version 2.14.1), distributed by the NASA High-Energy Astrophysics Archive Research Center (HEASARC) within the HEASOFT package, version 6.24. Instrumental responses were calculated based on the \nustar calibration database (CALDB), version 20180925. We cleaned and filtered raw event data for South Atlantic Anomaly (SAA) passages using the \texttt{nupipeline} module. We extracted source and background spectra from the calibrated and cleaned event files using the \texttt{nuproducts} module. Detailed information on these data reduction procedures can be found in the \nustar Data Analysis Software Guide \citep{nustardas}. A circular extraction radius of 30\arcsec\ was used for both the source and background regions. We extracted the background spectrum from source-free regions of the image on the same detector chip as the source, away from the outer edges of the field of view, which have systematically higher background. In order to maximise the available \nustar exposure, we extracted the `spacecraft science' mode 6 data, in addition to the standard `science' mode 1 data, following the method outlined in \citet{walton-2016}. We coadded the mode 1 and mode 6 spectra for each respective observation using the HEASOFT task \texttt{addspec}. We rebinned the spectral files using the HEASOFT task \texttt{grppha} to give a minimum of 20 photon counts per bin. 

\subsection{Archival XMM-Newton}

In addition to the \nustar observations taken in the 3--79 keV band, we analyzed archival \xmm observations taken in May 2009. 

We performed reduction of the \xmm data using the \xmm Science Analysis System (SAS, version 16.1.0), following the standard prescription outlined in the \xmm ABC online guide.\footnote{https://heasarc.gsfc.nasa.gov/docs/xmm/abc/} Calibrated, cleaned event files were created from the raw data files using the SAS commands \texttt{epchain} for the EPIC-pn detector \citep{xmm-pn} and \texttt{emchain} for the two EPIC-MOS detectors \citep{xmm-mos}. As recommended, we only extracted single and double pixel events for EPIC-pn and single to quadruple pixel events for EPIC-MOS. We excluded intervals of high background flux in the first $\simeq\!1/4$ of the observation, resulting in total source exposure of 8.4, 11.0, and 12.8 ks for the pn, MOS1 and MOS2 detectors, respectively. Source spectra were extracted from the cleaned event files using the SAS task \texttt{xmmselect} from a circular aperture with a radius of 40\arcsec\ centered on the source. Background spectra were extracted from a circular aperture with radius of 80\arcsec\ placed near the source on the same chip. Instrumental response files were generated for each of the detectors using the SAS tasks \texttt{rmfgen} and \texttt{arfgen}.

\subsection{Archival Swift}

In addition to the \nustar and \xmm spectra, we also analyzed \bat spectra collected over the first 70 months of observation \citep{swift-survey}, covering the 14--195 keV band. The \bat data reduction procedure is detailed in \citet{bat-105-month}. We do not include \xrt data in our broadband spectral modeling due to the poor signal-to-noise ratio of available \xrt data, which has $<$ 200 total photon counts. We separately analyzed archival \xrt data taken in 2005 to check for \nh variability. For this analysis, we reduced the \xrt data using the \texttt{XRTPIPELINE}, following the standard procedures detailed in \citet{evans-2009}.  


\section{X-ray Spectral Modeling}\label{sec:modeling}	

We performed joint spectral modeling of the broadband \nustar, \xmm, and \bat data using XSPEC v12.8.2 \citep{xspec}. We use $\chi^{2}$ statistics for all model fitting and error estimation and we quote uncertainties at the 90\% confidence level. We adopt cross sections from \citet{vern} and solar abundances from \citet{wilm}. We account for variability between different epochs by including cross-normalization factors between different datasets \citep[e.g.,][]{mislav-borus-2018}. In all our modeling, we include a Galactic absorption component with a fixed column density of \NHgal $= 1.62\times10^{21}$ cm$^{-2}$ \citep{galactic-nh}. We did not include \nustar data from the 2016 observation above 30 keV, as the source becomes background dominated above this energy, due to higher background levels relative to the 2017 observation. We model \nustar data from 2017 in the 3--79 keV band, and archival \xmm data in the 0.4--10 keV band.   

\begin{figure}[t]
 	\centering  
 	\begin{subfigure}{}
 	\hspace{-20pt} 
 		\includegraphics[width=0.5\textwidth]{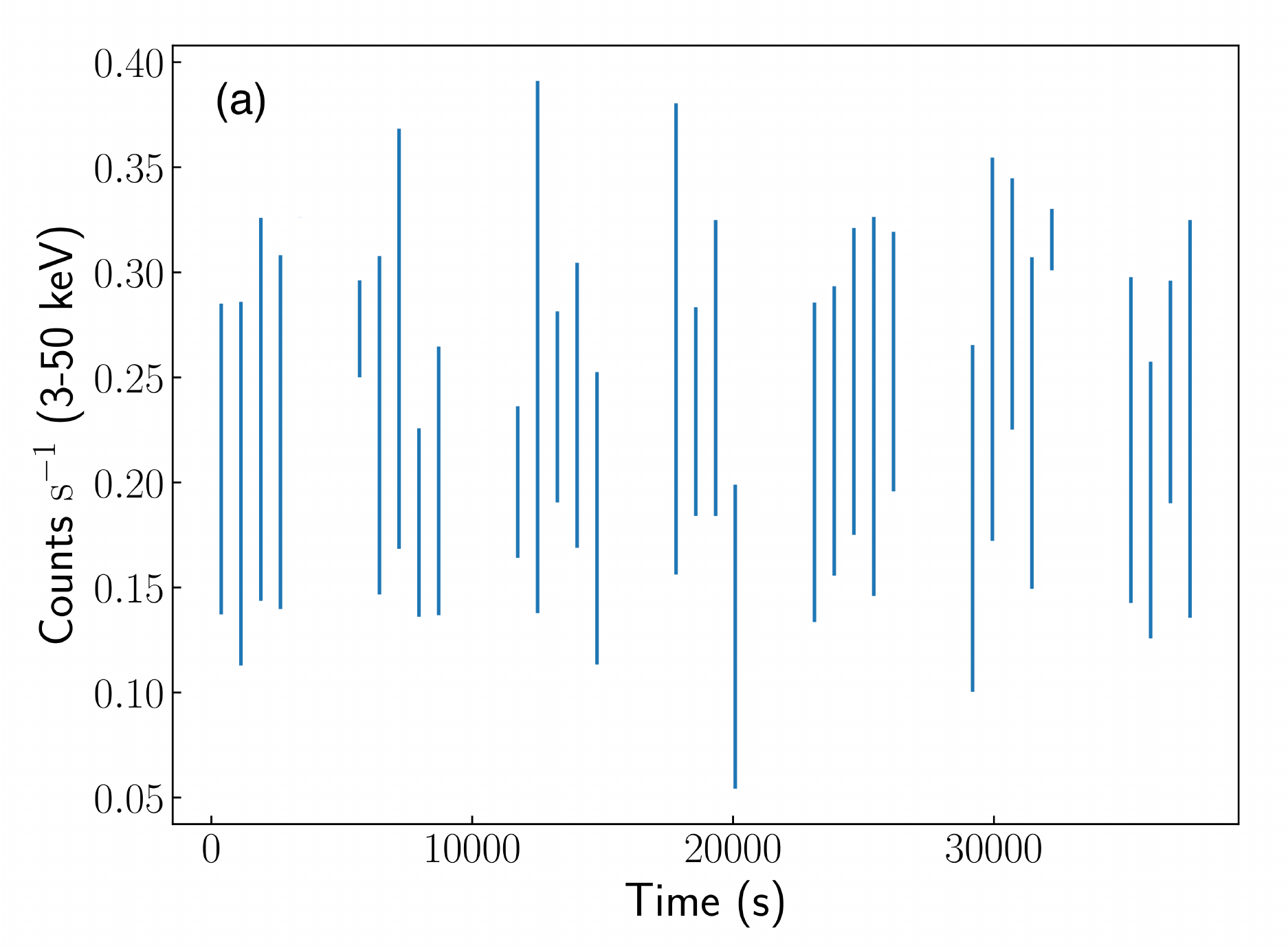}
 	\end{subfigure}%
 	\begin{subfigure}{}
 	\hspace{-20pt} 
 		\includegraphics[width=0.5\textwidth]{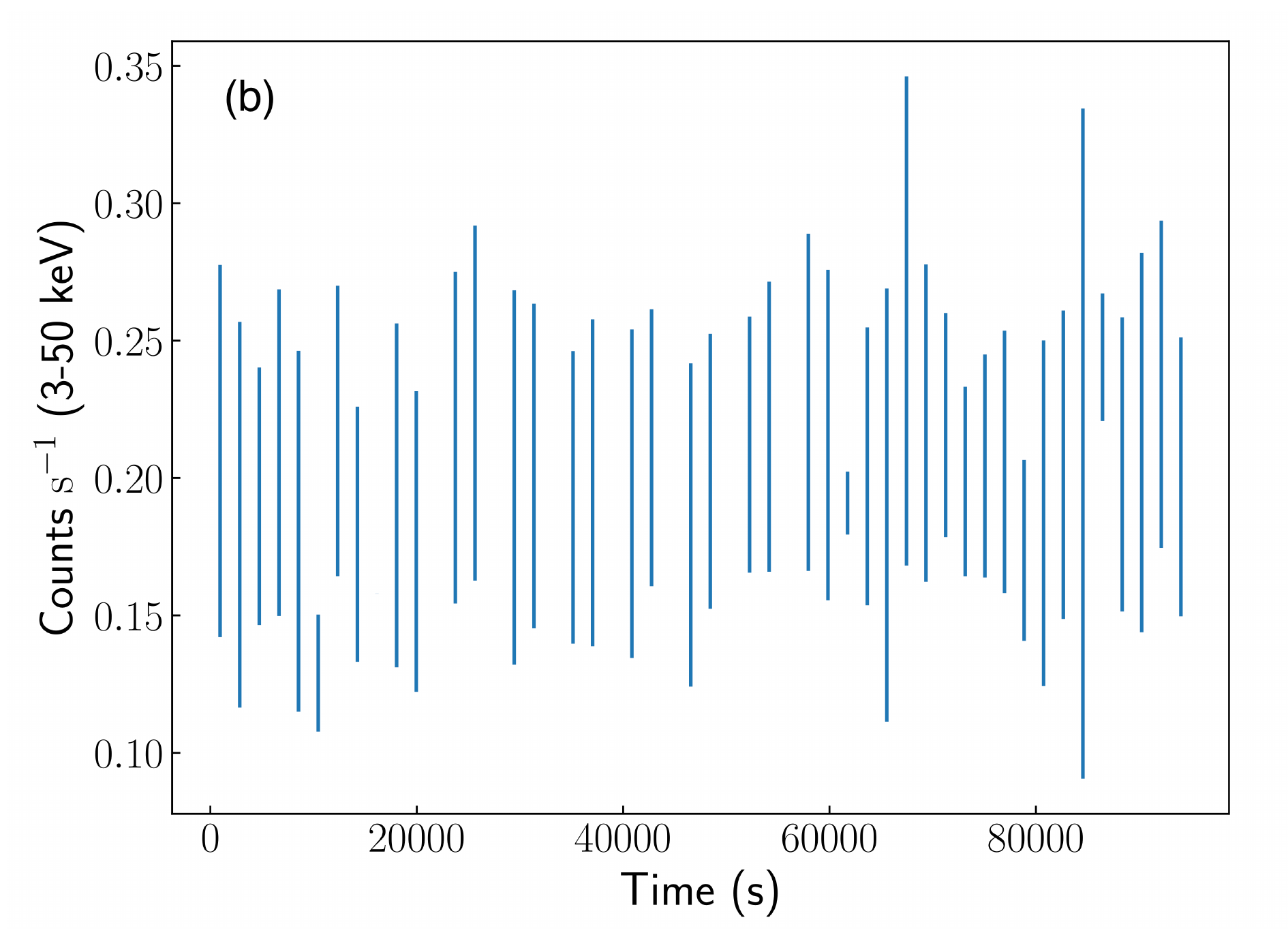}
 	\end{subfigure}%
 	\vspace{-10pt}
 	\caption{\nustar light curves of 2MASX J193013.80+341049.5 in the 3--50 keV band, where (a) is the light curve of the $\sim$ 20 ks observation taken in 2016 and (b) the deeper $\sim$ 50 ks observation taken in 2017. Gaps in data are due to Earth occultation events. There is no evidence for strong X-ray variability in this source.}
 	\label{fig:lightcurves}	
 \end{figure}

Analysis of the \nustar light curves for both observations in the 3--50 keV band (shown in Figure~\ref{fig:lightcurves}) show little evidence for strong variability in either the 2016 or 2017 observations, justifying the use of time-averaged spectra. Flux levels in the 2--10 keV band were also found to be similar between the two observations. \citet{hogg-2012} performed a detailed light curve analysis of the archival \xmm observation and found no significant variability in the \xmm EPIC-pn and MOS light curves.

 \begin{figure}[t]
	\hspace{-35pt}
	\includegraphics[width=0.57\textwidth]{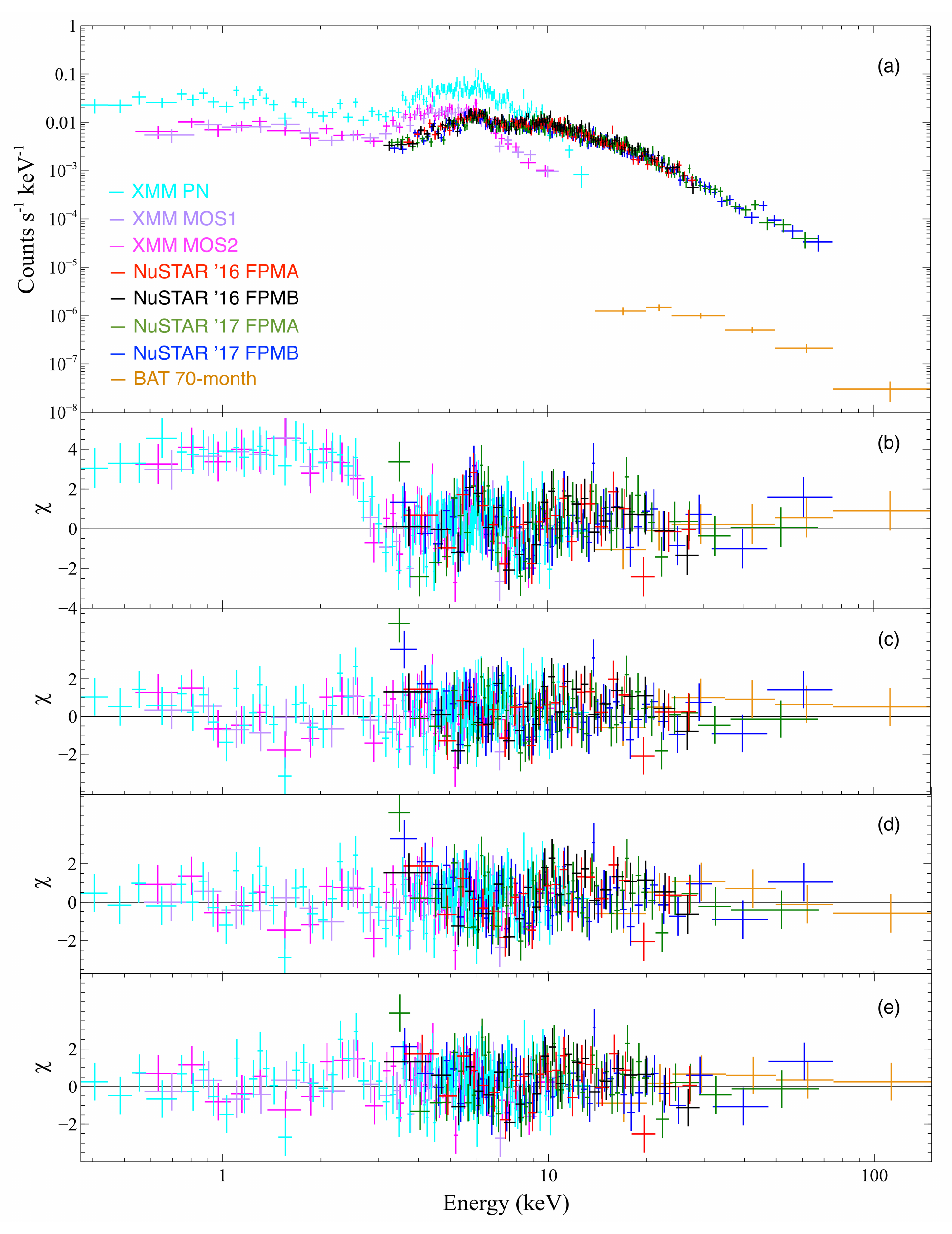}
	\caption{Broadband X-ray spectrum of 2MASX J19301380$+$3410495 (a), alongside fit residuals for (b) an absorbed cutoff power law model ($\chi^{2}$/dof = 1409/811), (c) absorbed cutoff power law with an Fe \ka line, scattered power law, and a reflection component incorporated using the \texttt{pexrav} model ($\chi^{2}$/dof = 826/798), (d) absorbed cutoff power law with a variable iron abundance absorber and reflection modeled with \texttt{pexmon} ($\chi^{2}$/dof = 816/798), and (e) \texttt{borus} model ($\chi^{2}$/dof = 808/795).}
	\vspace{5pt}
	\label{fig:specresiduals}	
\end{figure}

In our spectral modeling, we simultaneously fit the \xmm (EPIC-pn and EPIC-MOS), \nustar (FPMA and FPMB), and \bat spectra, covering a total energy range from 0.4 to 150 keV. We begin our analysis by fitting a simple absorbed cutoff power law model to the broadband data. As shown in Figure~\ref{fig:specresiduals}, an Fe K$\alpha$ emission line near 6.4 keV is evident in addition to excess soft emission below 3 keV. 

We construct models consisting of both phenomenological reflection models and physically-motivated torus models. All models include a photoelectric neutral absorber, power law continuum, and a scattered power law continuum. We leave the line-of-sight column density \nh free between different epochs. We list the applied models in XSPEC notation as follows:

\begin{enumerate}
	
	\item \texttt{phabs $\times$ (zphabs $\times$ cabs $\times$ cutoffpl $+$ zgauss $+$ pexrav $+$ constant $\times$ cutoffpl)}: Models an absorbed cutoff power law with a Gaussian Fe K$\alpha$ line, a cold Compton reflection component and a scattered power law component. \texttt{phabs} models Galactic absorption while \texttt{zphabs} models host galaxy absorption. \texttt{pexrav} \citep{pexrav} models reflection off a slab of infinite extent and optical depth covering between 0 and 2$\pi$ steradians of the sky relative to the illuminating source, corresponding to $R$ between 0 and 1. 
	
	\item \texttt{phabs $\times$ (cabs $\times$ TBFeo $\times$ cutoffpl $+$ pexmon $+$ constant $\times$ cutoffpl)}: Similar to Model 1, with the exception that \texttt{zphabs} is replaced with \texttt{TBFeo}, which models absorption with a variable iron abundance; the \texttt{zgauss} and \texttt{pexrav} components are replaced with \texttt{pexmon}, which is a slab reflection model that self consistently models the Fe \& Ni K complexes and Compton reflection hump assuming a semi-infinite plane geometry. 
	
	\item \texttt{phabs $\times$ (cabs $\times$ TBFeo $\times$ cutoffpl $+$ borus02 $+$ constant $\times$ cutoffpl)}: In the physically motivated \texttt{borus02} model \citep{mislav-borus-2018}, obscuring material is arranged in a toroidal structure around the central AGN, with a variable opening angle and torus column density that can be decoupled from the line-of-sight column density. This model provides self-consistent modeling of the fluorescent line emission and Compton reflection features.  
	
\end{enumerate}

\begin{table*}[t]
	\centering	
	\caption{Best-fit parameter values from modeling the broadband X-ray spectrum of 2MASX J193013.80+341049.5}
	\label{table:fitresults}
	\renewcommand{\arraystretch}{1.5}
	\begin{tabular*}{\textwidth}{@{\extracolsep{\fill} }l c c c c c}
	\noalign{\smallskip} \hline \hline \noalign{\smallskip}
	Model component & Parameter & & Model 1 & Model 2 & Model 3 \\ \hline 
	\texttt{zphabs} & \nh (\nustar 2016) & [$10^{23}$ cm$^{-2}$] & 3.3$\pm0.3$ & - & - \\
	 & \nh (\nustar 2017) & [$10^{23}$ cm$^{-2}$] & 3.8$\pm0.3$ & - & - \\
	 & \nh (\xmm) & [$10^{23}$ cm$^{-2}$] & 2.6$^{+0.3}_{-0.2}$ & - & - \\
	\texttt{TBFeo} & $A_{\text{Fe}}$ & [solar] & - & 1.07$^{+0.25}_{-0.30}$ & 0.45$^{+0.06}_{-0.05}$ \\
	 & \nh (\nustar 2016) & [$10^{23}$ cm$^{-2}$] & - & 4.8$^{+0.6}_{-0.5}$ & 5.2$^{+0.5}_{-0.1}$ \\ 
	 & \nh (\nustar 2017) & [$10^{23}$ cm$^{-2}$] & - & 5.5$^{+0.7}_{-0.5}$ & 4.9$^{+0.5}_{-0.7}$ \\
	 & \nh (\xmm) & [$10^{23}$ cm$^{-2}$] & - & 3.9$^{+0.4}_{-0.3}$ & 4.7$^{+0.3}_{-0.4}$ \\
	\texttt{pexrav} & $\Gamma^{A}$ & & 1.35$^{+0.18}_{-0.15}$ & - & - \\ 
	 & \ecut & [keV] & 49.9$^{+19.0}_{-11.2}$ & - & - \\
	 & $R$ & & 0.26$^{+0.32}_{-0.23}$ & - & - \\
	 & $A_{\text{Fe}}$ & [solar] & 1 (fixed) & - & - \\
	 & Norm$^{B}$ & [$10^{-3}$] & 1.41$^{+0.59}_{-0.37}$ & - & - \\
	\texttt{pexmon} & $\Gamma^{A}$ & & - & 1.33$^{+0.21}_{-0.14}$ & - \\
	 & \ecut & [keV] & - & 49.9$^{+38.3}_{-13.2}$ & - \\
	 & $R$ & & - & 0.25$^{+0.23}_{-0.10}$ & - \\
     & Norm$^{B}$ & [$10^{-3}$] & - & 1.47$^{+0.54}_{-0.34}$ & - \\
    \texttt{borus02} & $\Gamma^{A}$ & & - & - & 1.73$^{+0.05}_{-0.27}$ \\
     & \ecut & [keV] & - & - & 71.9$^{+21.2}_{-41.0}$ \\
     & log $N_{\text{H,Tor}}$ & & - & - & 24.25$^{+0.02}_{-0.12}$ \\
	 & $C_{f}^{C}$ & [\%] & - & - & $>$ 83.9 \\
	 & Norm$^{B}$ & [$10^{-3}$] & - & - & 2.44$^{+0.63}_{-0.81}$ \\ \hline
	\chidof &  &  & 826/798 & 816/798 & 808/795 \\ \hline
	\end{tabular*}
	\raggedright{$^{A}$Continuum photon index. \\ $^{B}$power law normalisation in units of counts s$^{-1}$ keV$^{-1}$ at 1 keV. \\ $^{C}$Torus covering factor.}
	\vspace{5pt}
\end{table*} 

\begin{figure*}[t]
	\includegraphics[width=0.9\linewidth]{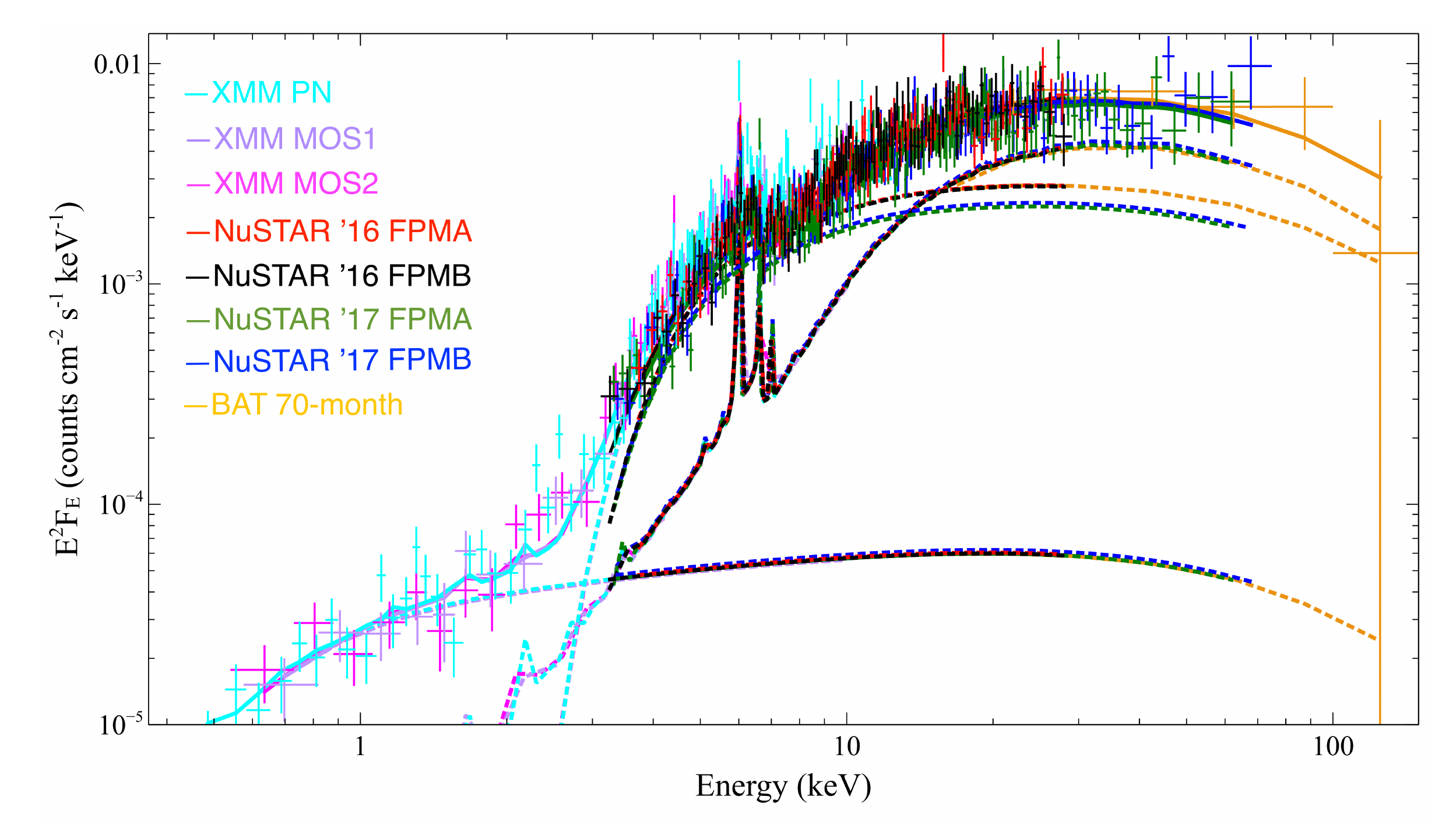}
	\caption{Broadband X-ray spectrum of 2MASX J193013.80+341049.5 unfolded through the \texttt{borus02} model (Model 3). Solid lines represent total model while dashed lines depict individual model components.}
	\vspace{5pt}
	\label{fig:boruseeuf}	
\end{figure*}

\begin{figure}[t]
	\hspace{-20pt}
	\includegraphics[width=0.53\textwidth]{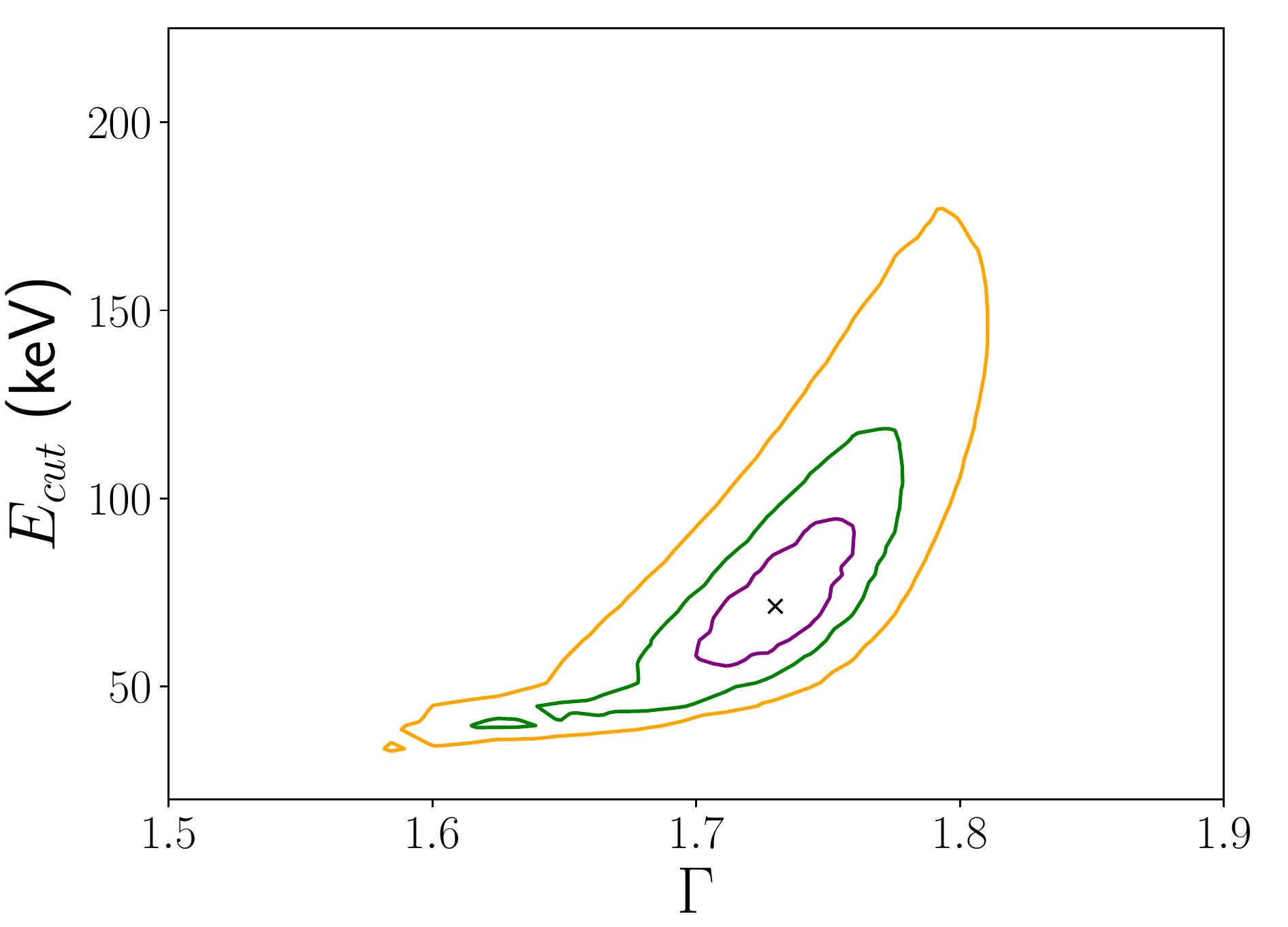}
	\caption{\ecut--$\Gamma$ contour plot from the \texttt{borus02} model fit (Model 3) to broadband X-ray data of 2MASX J193013.80+341049.5. The solid purple, green and yellow contours correspond to the 68, 90 and 99 \% confidence levels, respectively. The black cross represents the best-fit values of the parameters from applying the \texttt{borus02} model.}
	\label{fig:boruscontour}	
\end{figure} 
	  
For Model 1, we set iron and light element abundances to solar values and fix the inclination angle of the plane of reflecting material at the default value of cos $\theta = 0.45$. We tie the photon index and normalization of the reflected power law to that of the incident power law. We fix the energy of the Fe K$\alpha$ line at 6.4 keV. As shown in Figure~\ref{fig:specresiduals} (c), this model provides a considerable improvement in the fit compared to an absorbed cutoff power law model, with $\chi^{2}$/dof = 826/798. We find evidence of moderate reflection, with the reflection parameter $R$ constrained to 0.26$^{+0.32}_{-0.23}$. The cutoff energy is found to be strikingly low; \ecut = 50$^{+19}_{-11}$ keV. We find a fairly hard photon index ($\Gamma = 1.35^{+0.18}_{-0.15}$), which may partly account for the low cutoff energy obtained, due to degeneracy between \ecut and $\Gamma$, as evident in Figure~\ref{fig:boruscontour}. The line-of-sight column density exceeds $10^{23}$ cm$^{-2}$ in all epochs of observation, confirming this source to be highly X-ray absorbed. The value of \nh was similar between the \nustar observations, at $\sim (3-4)\times10^{23}$ cm$^{-2}$. For the archival \xmm observation, \nh was slightly lower at $\sim 2.6\times10^{23}$ cm$^{-2}$, but still consistent with high X-ray obscuration, and with the values reported by \citet{hogg-2012}. We also examined archival \xrt spectra taken in December 2005 and found high levels of absorption, with \nh found to be $5.3^{+5.5}_{-2.3}\times10^{23}$ cm$^{-2}$ from fitting Model 1 to the data. 

We find a fairly low equivalent width of the Fe K$\alpha$ line, at $\approx$ 80 eV for the \xmm observation and $\approx$ 90 eV for the \nustar observation. This motivates the exploration of alternative models with a variable iron abundance. In Model 2, we allow the iron abundance to vary by replacing the photoelectric absorption component \texttt{zphabs} with an absorber with variable iron abundance (\texttt{TBFeo}). We also replace the \texttt{pexrav} reflection model and \texttt{zgauss} line component with the phenomenological \texttt{pexmon} model, which self-consistently models both the reflection continuum and the Fe K$\alpha$ line. We tie the iron abundance parameter of the \texttt{pexmon} model to that of \texttt{TBFeo}. Model 2 yields an improvement to the fit compared to Model 1, with $\chi^{2}$/dof = 816/798. However, we find that the iron abundance is consistent with the Solar value when left as a free parameter. The reduction in $\chi^{2}$/dof compared to Model 1 is likely attributed to differences between the two models, such as the inclusion of additional emission lines in Model 2.  Furthermore, Model 2 produces slightly higher \nh values for a given epoch (see Table~\ref{table:fitresults}).

The phenomenological models used to account for reflection assume a simplistic slab geometry of the reflector with infinite extent and optical depth. While such models provide a convenient, simplified picture of the AGN reprocessor, they are not a physically realistic description of the geometry of the circumnuclear material, which is thought to have a roughly toroidal shape. We thus construct a third, physically-motivated model in which the \texttt{pexmon} component is replaced by \texttt{borus02} \citep{mislav-borus-2018}. The \texttt{borus02} model self-consistently computes the absorbed and reprocessed emission for a torus geometry with a central illuminating X-ray source. It provides more flexibility in modeling the torus geometry compared to previously developed torus models such as \texttt{MYTorus} \citep{mytorus}, as it includes the opening angle, the high-energy cutoff, and the relative iron abundance as free parameters.
In contrast, \texttt{MYTorus} assumes a uniform density torus with a fixed opening angle of {60\textdegree}, Solar abundance of iron and a termination energy at 500 keV. Both models can emulate torus clumpiness by allowing the average column density through the torus to be independent of the line-of-sight column density.

With Model 3, we set the covering factor of the torus, $C_{f}$, equal to cos $i$, where $i$ is the viewing angle of the torus. We find the fit naturally converges close to $C_{f} =$ cos $i$ when both parameters are left free to vary. We tie the iron abundance parameter of \texttt{borus02} to that of \texttt{TBFeo}. We link the photon index and normalization of the \texttt{borus02} component to that of the incident power law. Model 3 provides the best fit to the data compared to Models 1 and 2, with $\chi^{2}$/dof = 808/795. Leaving the iron abundance free results in a sub-solar value ($A_{\text{Fe}} = 0.45^{+0.06}_{-0.05}$), consistent with the weak Fe K$\alpha$ line observed in this source.

Table~\ref{table:fitresults} summarizes our modeling results for some of the key best-fit parameters for each of the three spectral models applied in this work. We conclude from our broadband spectral fitting that our physically-motivated \texttt{borus02} model (Model 3) provides the best fit to the data and also gives a photon index that lies within the typical range observed for Seyferts. We show the full broadband unfolded X-ray spectrum for the \texttt{borus02} model fit in Figure~\ref{fig:boruseeuf}.

With the additional, deeper 50 ks \emph{NuSTAR} observations taken 2017, the high-energy cutoff of the X-ray continuum is confirmed to be constrained to values that are atypically low compared to the Seyfert population.  \citet{ricci-2017a} found that the median cutoff energy for the \bat sample of unobscured AGN is $E_{\rm cut} = 210\pm36$ keV. Figure~\ref{fig:boruscontour} shows the contour plot of the photon index against the high-energy cutoff from the \texttt{borus02} model fit. While there is some degree of degeneracy between these two parameters, the value of \ecut is constrained to low values that are below the median cutoff energy found by \citet{ricci-2017a}. Such a low coronal cutoff is unusual but not unprecedented, with recent detections reported of AGN with high-energy cutoffs within the \nustar band \citep[e.g.,][]{grs-low-ecut,kara-2017,nikita-2018}. 

Various explanations have been proposed for the origin of low temperature coronae. For sources accreting at high Eddington rates, Compton cooling may be enhanced due to the stronger radiation field present in comparison to lower Eddington ratio Seyferts \citep{kara-2017}. For AGN accreting well below the Eddington limit, low coronal temperatures may be attributed to a high optical depth within the corona, which results in more effective cooling due to multiple inverse Compton scatterings of seed photons from the accretion disk \citep{grs-low-ecut}. Low temperatures can also be achieved if the corona consists of a hybridized plasma, containing both thermal and non-thermal particles \citep[e.g.,][]{Zdziarski-1993,corona-heating-paper,fabian-2017}. In such a hybridized system, only a small fraction of non-thermal electrons with energies above 1 MeV are needed to result in runaway electron-positron pair production. The cooled pairs redistribute their available energy, thereby reducing the mean energy per particle and decreasing the coronal temperature. Such cooling would produce a hard non-thermal tail in the X-ray spectrum and an annihilation feature at 511 keV, both of which are currently undetectable with present X-ray instrumentation. In order to robustly test hybrid
plasma models, next-generation hard X-ray observatories with high sensitivity at
energies beyond 100 keV, such as the \emph{High-Energy X-ray Probe} (\emph{HEX-P}) \citep{madsen-2018}, will be essential.

\section{Multi-wavelength Analysis}\label{sec:multiwavelength}

\begin{table*}[t]
	\centering	
	\caption{Observation details of the optical spectra of 2MASX J193013.80+341049.5 considered in this work.}
	\renewcommand{\arraystretch}{1.0}
	\label{table:opticalobs}
	\begin{tabular*}{\textwidth}{@{\extracolsep{\fill} }l c c c c c c c}
		
		\noalign{\smallskip} \hline \hline \noalign{\smallskip}
		
		Observation Date & Telescope & Instrument & Exposure time & Slit Size & Slit Size & Spectral Resolution (FWHM) \\
		& & & (s) & (\si{\arcsec}) & (kpc) & (\si{\angstrom})  \\ \hline
		2006-05-08 & Orzale & BFOSC & 3600 & 2 & 2.4 & 13   \\
		2007-06-16 & KPNO & Goldcam & 3600 & 2 & 2.4 & 3.3   \\
		2016-10-02 & Palomar & DBSP & 300 & 1.5 & 1.8 & 4.2, 5.9  \\
		2018-08-11 & Palomar & DBSP & 300 & 2 & 2.4 & 4.8, 7.4   \\
		2019-08-28 & Palomar & DBSP & 900 & 1.5 & 1.8 & 4.4, 5.8  
		\\ \hline	
	\end{tabular*}
\end{table*}

\begin{table*}[t]
	\centering	
	\caption{Flux details and selected line-width properties of the optical spectra of 2MASX J193013.80+341049.5 considered in this work.}
	\renewcommand{\arraystretch}{1.0}
	\label{table:opticalspec}
	\begin{tabular*}{\textwidth}{@{\extracolsep{\fill} }l c c c c c c}
		
		\noalign{\smallskip} \hline \hline \noalign{\smallskip}
		
		Observation Date & Continuum flux$^{A}$ & H$\alpha$ flux & H$\alpha$ FWHM & H$\beta$ flux & H$\beta$ FWHM \\
		& (erg s$^{-1}$ cm$^{-2}$ \si{\angstrom}$^{-1}$) & ($10^{-17}$ erg s$^{-1}$ cm$^{-2}$) & (km s$^{-1}$) & ($10^{-17}$ erg s$^{-1}$ cm$^{-2}$) & (km s$^{-1}$) \\ \hline
		2006-05-08 & 160 & 60501 & 6948 & 8580 & 3577 \\
		2007-06-16 & 77 & 37097 & 7053 & 4363 & 3373 \\
		2016-10-02 & 78 & 41239 & 6720 & 3189 & 3639 \\
		2018-08-11 & 48 & 33022 & 6697 & 1822 & 3610 \\
		2019-08-28 & 58 & 30915 & 7227 & 2657 & 4385
		\\ \hline	
	\end{tabular*}
	\raggedright{$^{A}$ Continuum flux density estimated in the 4760--4780 \si{\angstrom} band.}
\end{table*}

\begin{figure*}[t]
	\centering
	\hspace{-25pt}
	\includegraphics[width=0.8\linewidth]{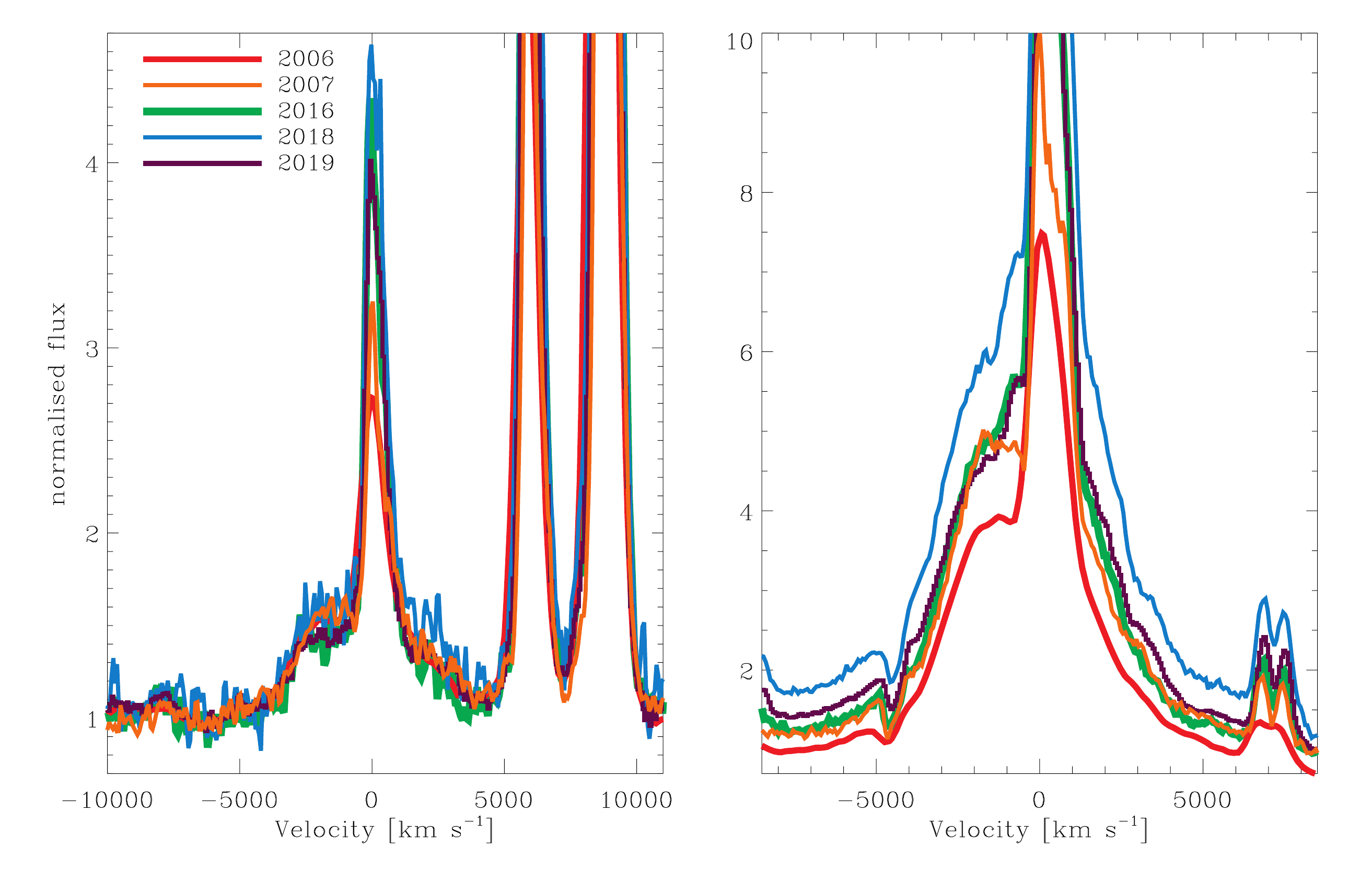}
	\caption{Optical spectra of 2MASX J193013.80+341049.5 taken at Orzale (2006), KPNO (2007), and Palomar Observatory (2016, 2018, 2019). The fluxes have been normalized by the emission in a line free region (4760 - 4780 \AA). Left and right panels show the velocity centered on the H$\beta$ and H$\alpha$ line profiles respectively.}
	\vspace{5pt}
	\label{fig:opticalspec}	
\end{figure*}

From broadband X-ray spectral modeling, we confirm that 2MASX J193013.80+341049.5 exhibits high levels of X-ray obscuration, with \nh exceeding $10^{23}$ cm$^{-2}$ in both the \nustar and archival \xmm observations. The X-ray spectral features of this source are characteristic of a classic obscured, Type 2 AGN. The high X-ray absorption present is thus in clear conflict with the optical classification of this source as a Type 1 AGN. We explore properties of this enigmatic source at other wavelengths to investigate whether there are other unusual features possibly linked to the high X-ray obscuration but comparatively lower optical obscuration present in this source. We also discuss possible mechanisms for producing such mismatches between X-ray and optical classifications.

\subsection{Optical Spectra}

2MASX J193013.80+341049.5 has consistently been classified as a Type 1 AGN from optical spectra taken over several epochs \citep[e.g.,][]{landi-2007,trippe-2011}, which show clear broad components to the H$\alpha$ and H$\beta$ lines. We also obtained recent optical spectra using the Double Spectrograph (DBSP) instrument on the 200-inch Hale telescope at Palomar Observatory. Observations were performed in 2016 October, 2018 August, and 2019 August. We compare the Palomar spectra with those taken at Orzale in 2006 and KPNO in 2007 \citep{koss-2017}. Full observation details of the optical spectra examined in this work are presented in Table~\ref{table:opticalobs}. Broad H$\alpha$ and H$\beta$ lines are present in all epochs, consistent with a Type 1 optical classification (Figure~\ref{fig:opticalspec}). We also observed a nearby galaxy companion during the UT 2019 August 28 observation, 2MASS J193015.12+34111.18, which is 27.3\arcsec\ to the northeast of the AGN (33.4 kpc).  The galaxy is an absorption line system at a close redshift ($z=0.063$ based on the Ca H+K absorption lines) to the primary AGN galaxy.

For emission line measurements, we follow the procedure used in the OSSY database \citep{oh-2011} and its broad-line prescription \citep{oh-2015}. We apply stellar templates \citep{bruzual-charlot-2003,sanchez-2006} and emission-line fitting in a rest-frame ranging from 3780 \si{\angstrom} to 7580 \si{\angstrom}.  Using the broad H$\alpha$ flux and line-widths from the 2016 observation, we derive a black hole mass of $\log (M_{\rm BH}/M_\odot) = 8.3$ using the virial approximation of \citet{green-ho-2007}. The flux of the  broad H$\beta$ line, broad H$\alpha$ line, and continuum (4760--4780 \AA) are observed to decrease over the period 2006-2019 (Table~\ref{table:opticalspec}) by more than a factor of 2. Errors are dominated by flux calibration of the order 10--20\%.  The H$\beta$ line FWHM remains roughly constant across observations. It is unlikely that the mismatch in X-ray and optical classifications is due to short timescale variability of the line-of-sight absorption, as \nh remains consistently high in all epochs of X-ray observations. Furthermore, \citet{hogg-2012} note little variability in optical spectra taken three months apart \citep{landi-2007,winter-2010}.


\subsection{X-ray to Mid-IR Relation}

\begin{figure}[t]
	\hspace{-20pt}
	\includegraphics[width=0.53\textwidth]{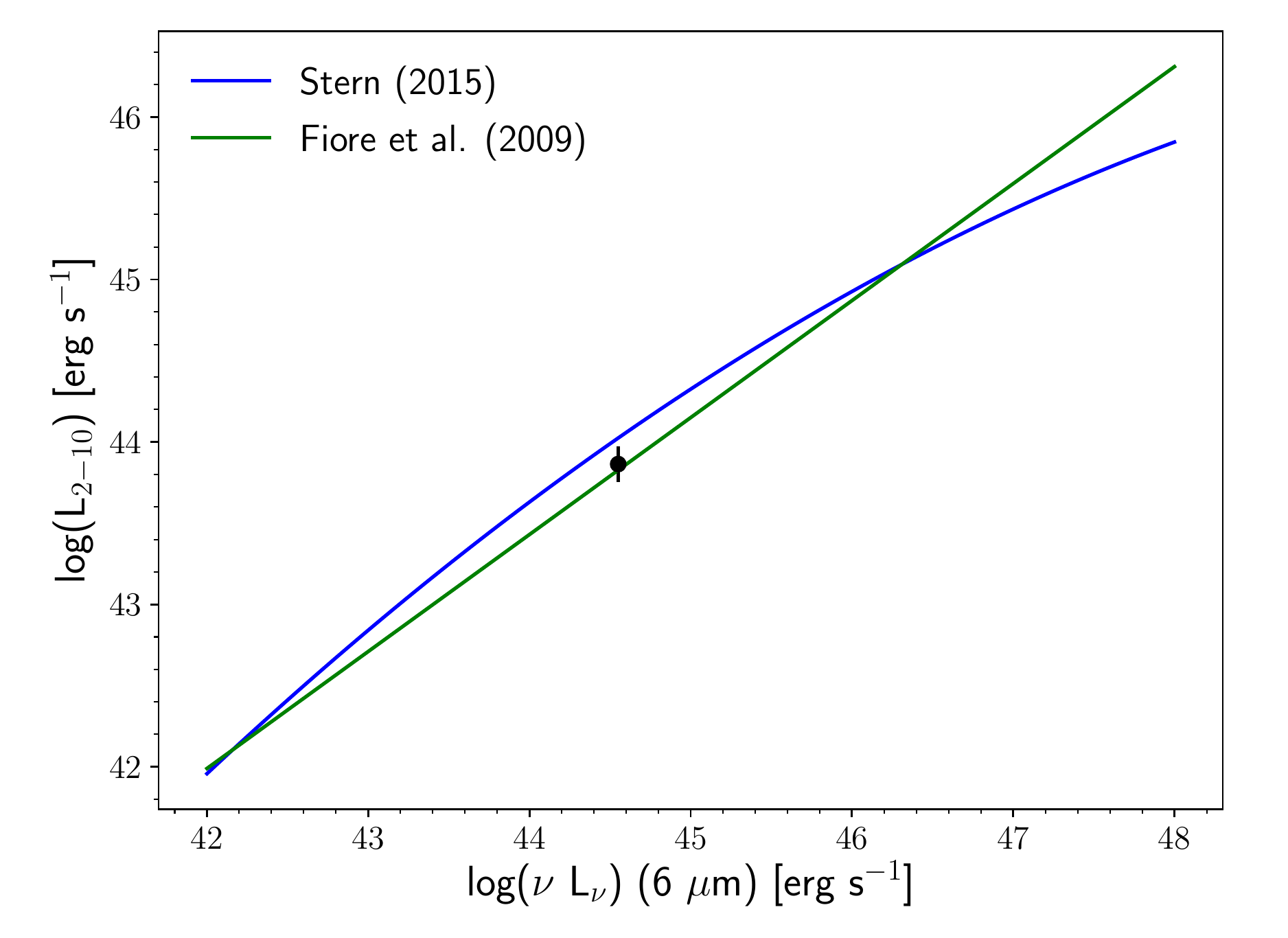}
	\caption{Rest-frame 2--10 keV X-ray luminosity against rest-frame 6 $\mu$m mid-infrared luminosity for 2MASX J193013.80+341049.5 (black point) along with published relations from \citet{fiore-2009} and \citet{stern-2015}.}
	\label{fig:midIRXray}	
\end{figure}

It is well-established that the X-ray and mid-IR emission from AGN are correlated \citep[e.g.,][]{lutz-2004,gandhi-2009,stern-2015}. We investigate whether 2MASX J193013.80+341049.5 has mid-IR properties that are consistent with the observed correlation between mid-IR and X-ray luminosity. We estimate the mid-IR luminosity at 6 $\mu$m by constructing the broadband SED for 2MASX J193013.80+341049.5 using publicly available data from the \emph{Vizier} catalog. The data obtained from \emph{Vizier} includes measurements from catalogs such as 2MASS, {\it Gaia}, PanSTARRS, {\it AKARI}, {\it WISE} and {\it XMM}-UVOT, covering a total wavelength range from 0.3--100 $\mu$m. We linearly interpolate between 3.75 and 15 $\mu$m to determine the rest-frame 6 $\mu$m flux density. We convert from specific flux to luminosity using a luminosity distance of 268 Mpc obtained from the NASA/IPAC Extragalactic Database (NED). We calculate the X-ray luminosity from the unabsorbed flux in the 2--10 keV band, using the 2017 \nustar data. We also estimated the predicted 2--10 keV luminosity based on the H$\alpha$ luminosity measured from the the 2016 optical spectra, using the observed relation between intrinsic 2--10 keV luminosity and H$\alpha$ luminosity \citep{panessa-2006}. We found the predicted X-ray luminosity to be consistent with the intrinsic luminosity observed from the \nustar data. 

Figure~\ref{fig:midIRXray} shows the rest-frame 2--10 keV luminosity against the rest-frame 6 $\mu$m luminosity for 2MASX J193013.80+341049.5, along with correlations reported in the literature. \citet{fiore-2009} report a mid-IR to X-ray correlation based on samples of X-ray-selected Type 1 AGN in the COSMOS and CDFS fields. The relation presented in \citet{stern-2015} was obtained by including a sample of the most luminous quasars from the Sloan Digital Sky Survey, and is appropriate for AGN across a large luminosity range, spanning from the Seyfert through to the quasar regime. The location of our source is consistent with both these published relations, indicating that 2MASX J193013.80+341049.5 is not atypical in terms of its mid-IR/X-ray luminosity ratio.      

\subsection{Broadband SED and \aox}

\begin{figure}[t]
	\hspace{-30pt}
	\includegraphics[width=0.55\textwidth]{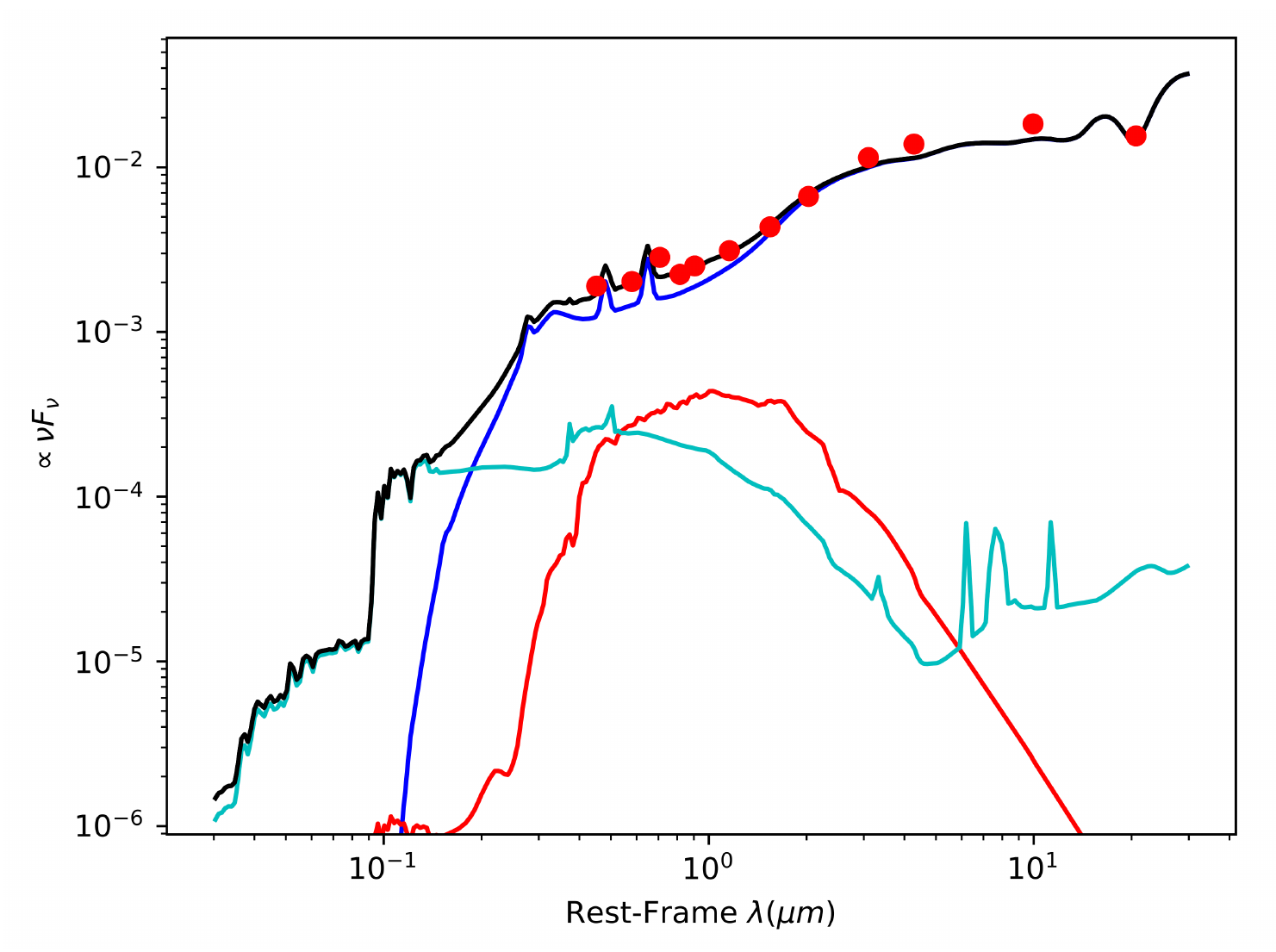}
	\caption{Spectral energy distribution (SED) of 2MASX J193013.80+341049.5. The best-fit SED model (black line) consists of an AGN component (blue line), a young stellar population (cyan line), and an old stellar population (red line). Solid red points are observed flux densities used for the SED fit, obtained from publicly available data from the \emph{Vizier} catalog and PanSTARRS DR2.}
	\label{fig:sedfit}	
\end{figure}

We construct the broadband SED for 2MASX J193013.80+341049.5 using publicly available flux densities from the \emph{Vizier} catalog and PanSTARRS DR2. We perform SED fitting in order to investigate the relationship between UV and X-ray luminosity for this source, parameterized by the \aox spectral slope, defined as:

\begin{center}
	\begin{equation*} 
	\alpha_{\text{ox}} = -0.384\times \text{Log}[L_{2
	\text{keV}}/L_{2500}]
	\end{equation*}
\end{center}

\noindent where $L_{2 \text{keV}}$ and $L_{2500}$ are the monochromatic luminosities at 2 keV and 2500 \si{\angstrom}, respectively. We determine the monochromatic luminosity at 2500 \si{\angstrom}, corrected for dust-reddening, via SED template fitting. We model the SED of 2MASX J193013.80+341049.5 in the 0.03--30 $\mu$m range using the algorithm and empirical templates of \citet{assef-2010}. The models consist of a linear combination of a dust-reddened AGN template and three empirical galaxy templates, corresponding to E, Sbc, and Im type galaxies. We do not include UV data in our SED fitting due to the uncertainty
in the UV extinction correction. Figure~\ref{fig:sedfit} presents the best-fit SED model of 2MASX J193013.80+341049.5 along with individual model components. We linearly interpolate the best-fit, un-reddened AGN component to determine the intrinsic specific flux at 2500 \si{\angstrom}. We compute the monochromatic X-ray luminosity at 2 keV from modeling the broadband X-ray spectra, correcting for absorption.

\begin{figure}[t]
	\hspace{-20pt}
	\includegraphics[width=0.55\textwidth]{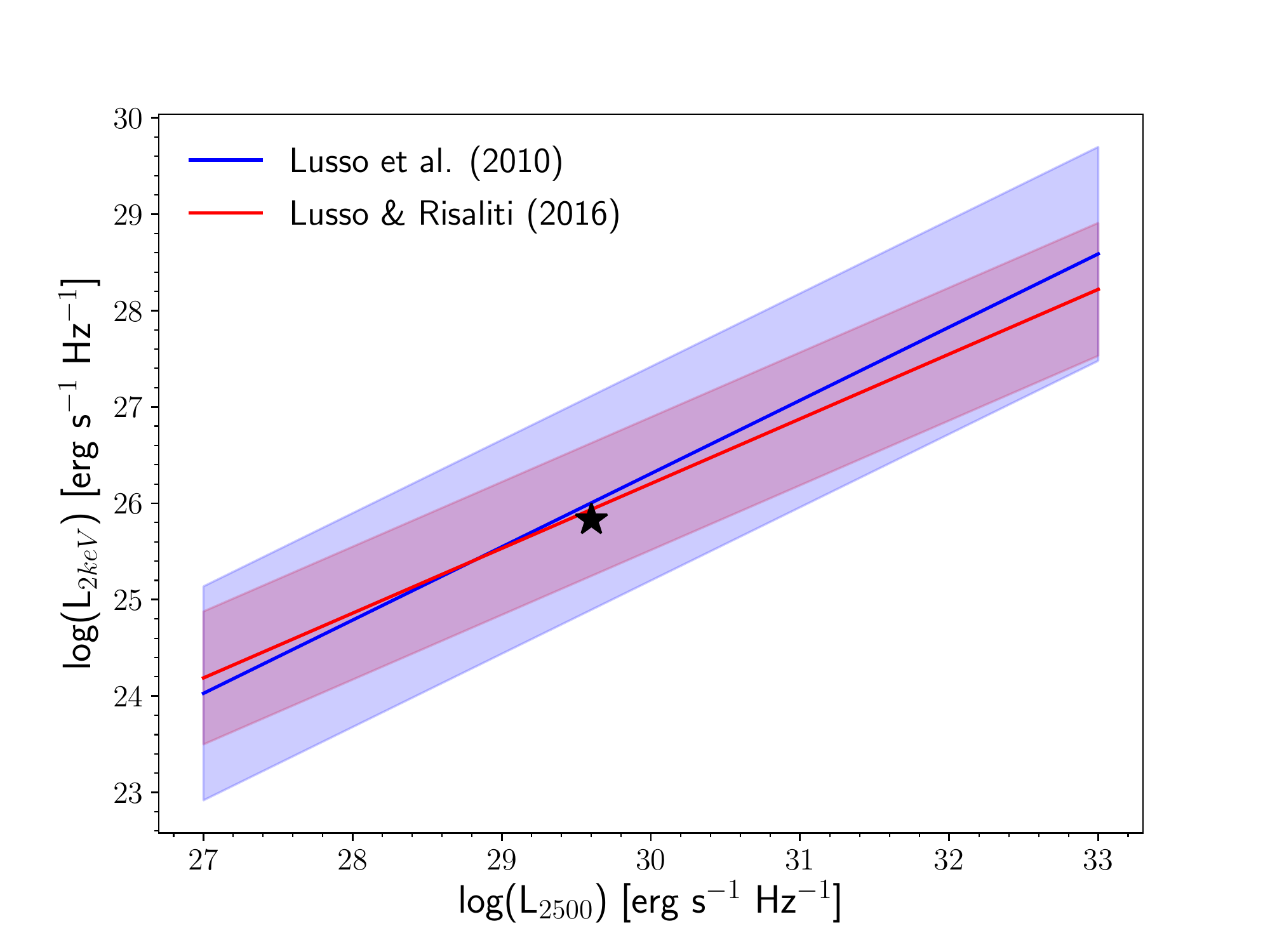}
	\caption{Rest-frame monochromatic 2 keV luminosity $L_{2 \text{keV}}$, against rest-frame 2500 \si{\angstrom} luminosity $L_{2500}$ for 2MASX J193013.80+341049.5 (black star). Also plotted are literature relations from \citet{lusso-2010} and \citet{lusso-2016}. The shaded regions represent the 3 $\sigma$ dispersion in the fitted relations.}
	\label{fig:L2500vsL2keV}	
\end{figure}

In Figure~\ref{fig:L2500vsL2keV} we present the monochromatic X-ray luminosity as a function of the monochromatic UV luminosity, along with fitted relations reported in the literature. \citet{lusso-2010} present a significant correlation between $L_{2 \text{keV}}$ and $L_{2500}$ based on a sample of 545 X-ray-selected Type 1 AGN from the XMM-COSMOS survey, spanning a wide range of redshifts and X-ray luminosities. In \citet{lusso-2016}, a tighter correlation is reported using a sample of 2685 quasars that have been optically selected with homogeneous SED and X-ray detections, with dust-reddened and gas-obscured sources excluded from the sample. 2MASX J193013.80+341049.5 is consistent with observed relations between $L_{2 \text{keV}}$ and $L_{2500}$ (Figure~\ref{fig:L2500vsL2keV}), indicating that this source is also not atypical in terms of its UV/X-ray luminosity ratio. We find \aox $\sim$ 1.4, which is within the typical range of \aox distributions, which covers 1.2 - 1.8 \citep{lusso-2010}.

The normal X-ray-to-optical ratio (parameterized by \aox) of 2MASX J193013.80+341049.5 indicates that it is unlikely that the Type 1 optical classification is due to scattering of BLR photons into our line-of-sight through a region of lower column density. If there were strong scattering present, a stronger X-ray-to-optical ratio would be expected due to suppression of optical emission. We further rule out a scattering scenario through our broadband SED fitting, where we allow for a second unobscured AGN component to account for the scattered light leaked from a primary obscured AGN component. The best-fit for such a model produces zero flux for the second AGN component, indicating no evidence for scattered/reflected light leaking into our line of sight. 


\subsection{Gas-to-Dust Ratio}

Some studies have reported Type 1 AGN with large X-ray column densities \citep[e.g.,][]{shimizu-2017}, however explanations
for the existence of such objects have varied widely. One explanation that is consistent with the unified model is that our line of sight grazes the edge of the obscuring torus where the cloud distribution is less dense, but still provides significant X-ray absorption due to the X-ray corona's small physical size compared to the BLR. This would result in a larger effective covering fraction of the compact corona in comparison with the more extended BLR. This scenario is supported from our X-ray modeling with \texttt{Borus}, where we found that a geometry in which the torus is viewed through the rim provided the best-fit to the data. 

If we model the torus to have a clumpy distribution \citep{krolik-1988}, then another possibility is that a clump has entered our line of sight during the X-ray observation, causing a temporary increase in the X-ray column density. However, the lack of variability in \nh between epochs of X-ray observations appear to disfavour such short timescale variability as the cause of the X-ray/optical mismatch (see also Section 4.1).

High levels of X-ray obscuration but lack of optical obscuration could also be explained by the presence of high-density, ionized gas outflows. For example, the broad-line radio galaxy 3C 445 is classified as Type 1 from broad H$\alpha$ and H$\beta$ lines in its optical spectrum, however \emph{Suzaku} and \emph{Chandra} observations show it to be heavily absorbed with $N_{\rm H}\sim10^{23}$ cm$^{-2}$. The soft X-ray spectrum of this source is dominated by ionized emission lines \citep{braito-2011}. A mechanism proposed  by \citet{braito-2011} is that the photo-ionized, outflowing absorber is associated with a disk wind that is clumpy in nature and located close to the central X-ray source. The clumpy distribution of the absorber enables visible sightlines to the BLR gas. High resolution X-ray grating observations are generally needed to identify emission lines or absorption features associated with ionized gas outflows. 


\vspace{10pt}
\begin{figure}[t]
	\hspace{-20pt}
	\includegraphics[width=0.55\textwidth]{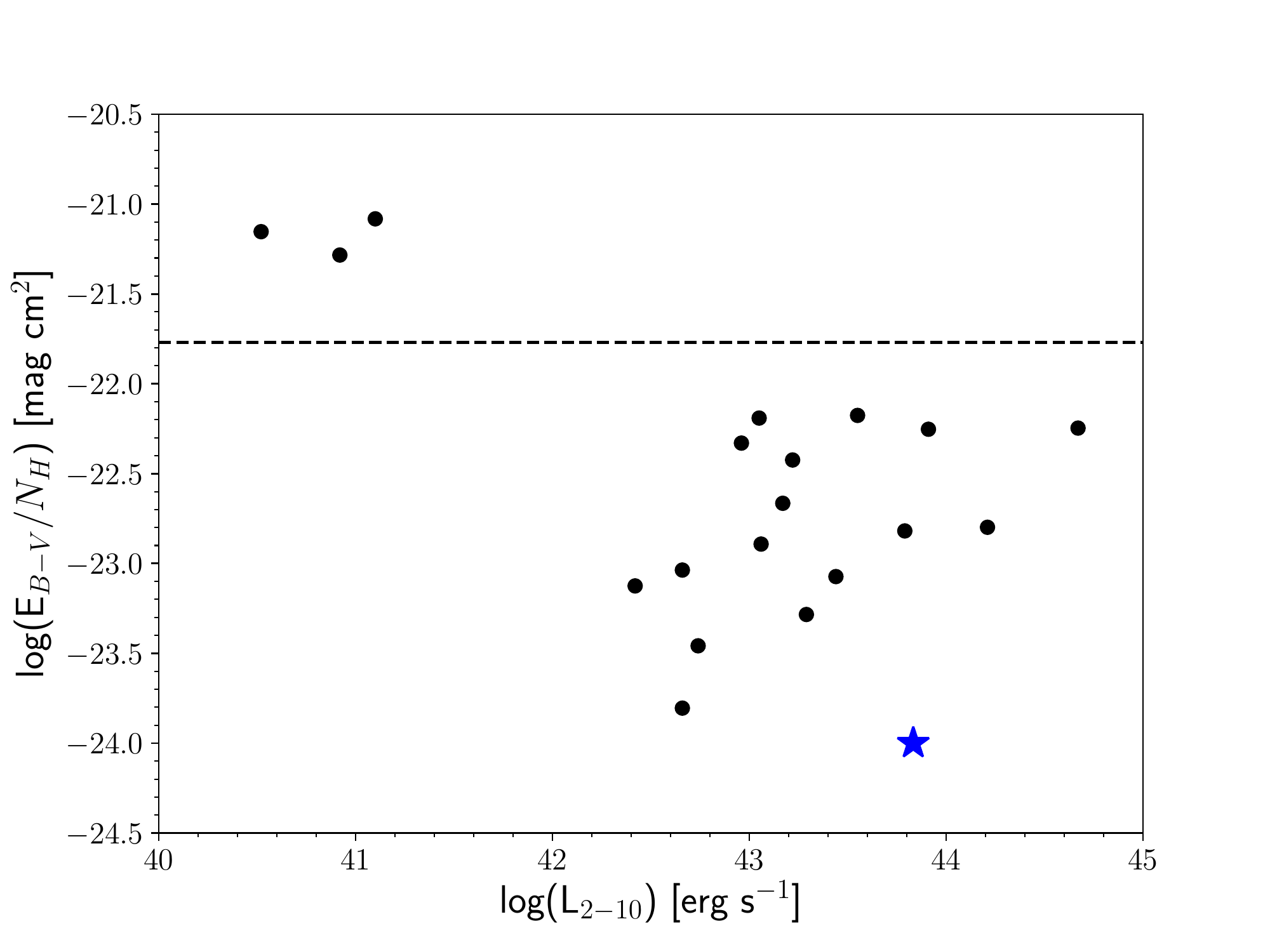}
	\caption{$E_{B-V}$/\nh ratio vs. intrinsic 2--10 keV X-ray luminosity for 2MASX J193013.80+341049.5 (blue star), and for sources from the sample of \citet{maiolino-2001} (black points). The reddening $E_{B-V}$ is estimated assuming a Galactic standard extinction curve. The column density \nh is determined from fitting X-ray spectra. Dashed black line represents the Galactic standard value of $E_{B-V}$/\nh.}
	\label{fig:EBV-NH}	
\end{figure}

Another possibility for the mismatch in X-ray and optical classifications for 2MASX J193013.80+341049.5 is a lower dust-to-gas ratio relative to the Galactic interstellar medium. \citet{maiolino-2001} investigated the ratio of reddening to the X-ray absorbing column density, $E_{B-V}$/\nh, for a diverse sample of AGN characterised by cold X-ray absorption. They found that $E_{B-V}$/\nh is lower than Galactic by factors ranging from $\sim$ 3--100 for most sources in their sample, assuming a standard Galactic extinction curve. In Figure~\ref{fig:EBV-NH}, we show $E_{B-V}$/\nh as a function of the intrinsic 2--10 keV luminosity for 2MASX J193013.80+341049.5 compared to the sample from \citet{maiolino-2001}. We find $E_{B-V} = 0.50\pm0.07$, determined from our broadband SED fitting. We observe that 2MASX J193013.80+341049.5 has an $E_{B-V}$/\nh that is significantly lower than the Galactic standard value by a factor of 170. This suggests that reduced dust absorption compared to the gaseous column density may likely explain the mismatch between the X-ray and optical classification of 2MASX J193013.80+341049.5. It is possible that the BLR itself provides extra X-ray obscuration and consists of neutral, dust-free gas that is an inner extension of the dusty molecular torus \citep{davies-2015}. If the BLR is the source of the X-ray absorption, \nh variability may be seen on relatively short timescales \citep[e.g.,][]{risaliti-2005}. One method to verify where the bulk of the absorbing gas resides is through resolving the width of the Fe K$\alpha$ line \citep[e.g.,][]{gandhi-2015}. Future X-ray missions such as \emph{Athena} \citep{athena-2015} will be able to measure the width of the Fe K$\alpha$ line with unprecedented spectral resolution.

\section{Summary}\label{sec:summary}

In this paper, we present both a broadband X-ray and a multi-wavelength analysis of the enigmatic X-ray obscured, but optically unobscured Type 1 AGN 2MASX J193013.80+341049.5. From joint modeling of \nustar and archival \xmm observations, we find 2MASX J193013.80+341049.5 to be strongly absorbed, with \nh $> 2\times10^{23}$ cm$^{-2}$. We also find the source to possess an atypically low coronal temperature compared to the Seyfert population (\ecut $\sim$ 72 keV). Some possible mechanisms to explain the low coronal temperature include a large optical depth of the corona, a strong radiation field or a hybrid pair-dominated plasma.  

We investigate possible scenarios to explain the mismatch between X-ray and optical classifications using techniques such as broadband SED modeling.  We find 2MASX J193013.80+341049.5 likely has a much lower dust-to-gas ratio relative to the Galactic ISM, with $E_{B-V}$/\nh lower than the Galactic standard by a factor of $\sim$ 170. This suggests that the X-ray/optical mismatch could be explained by the BLR itself providing the source of extra X-ray obscuration, and is composed of low-ionization, dust-free gas.   




\section{Acknowledgements}

We have made use of data from the \nustar mission, a project led by the California Institute of Technology, managed by the Jet Propulsion Laboratory, and funded by the National Aeronautics and Space Administration. We thank the \nustar Operations, Software and Calibration teams for support with the execution and analysis of these observations. This research has made use of the \nustar Data Analysis Software (NuSTARDAS) jointly developed by the ASI Science Data Center (ASDC, Italy) and the California Institute of Technology (USA). M.B. acknowledges support from the Black Hole Initiative at Harvard University, through a grant from the John Templeton Foundation. M.K. acknowledges support from NASA through ADAP award NNH16CT03C. D.J.W. acknowledges support from STFC in the form of an Ernest Rutherford fellowship. R.J.A was supported by FONDECYT grant number 1191124.

\facilities{\nustar, \emph{Swift}, \xmm, \emph{Palomar DBSP}}

\software{NuSTARDAS (v2.14.1), HEASOFT (v6.24), XMM SAS (v16.1.0), XSPEC (v12.8.2), Astropy, SciPy, NumPy, Matplotlib}

\vspace{15pt}

\bibliographystyle{apj}
\bibliography{BAT_Sy1_refs}

\begin{thebibliography}{}
\expandafter\ifx\csname natexlab\endcsname\relax\def\natexlab#1{#1}\fi

\bibitem[{Antonucci(1993)}]{antonucci}
Antonucci, R. 1993, \aap, 31, 473

\bibitem[{Arnaud(1996)}]{xspec}
Arnaud, K. 1996, in ASP Conf. Series, Vol. 101, Astronomical Data Analysis
  Software and Systems, ed. G.~H. {Jacoby} \& J.~{Barnes}, 17

\bibitem[{{Assef} {et~al.}(2010){Assef}, {Kochanek}, {Brodwin}, {Cool},
  {Forman}, {Gonzalez}, {Hickox}, {Jones}, {Le Floc'h}, {Moustakas}, {Murray},
  \& {Stern}}]{assef-2010}
{Assef}, R.~J., {Kochanek}, C.~S., {Brodwin}, M., {et~al.} 2010, \apj, 713, 970

\bibitem[{{Balokovi{\'c}} {et~al.}(2018){Balokovi{\'c}}, {Brightman},
  {Harrison}, {Comastri}, {Ricci}, {Buchner}, {Gandhi}, {Farrah}, \&
  {Stern}}]{mislav-borus-2018}
{Balokovi{\'c}}, M., {Brightman}, M., {Harrison}, F.~A., {et~al.} 2018, \apj,
  854, 42

\bibitem[{{Barcons} {et~al.}(2015){Barcons}, {Nandra}, {Barret}, {den Herder},
  {Fabian}, {Piro}, {Watson}, \& {the Athena Team}}]{athena-2015}
{Barcons}, X., {Nandra}, K., {Barret}, D., {et~al.} 2015, in Journal of Physics
  Conference Series, Vol. 610, Journal of Physics Conference Series, 012008

\bibitem[{{Baumgartner} {et~al.}(2013){Baumgartner}, {Tueller}, {Markwardt},
  {Skinner}, {Barthelmy}, {Mushotzky}, {Evans}, \& {Gehrels}}]{swift-survey}
{Baumgartner}, W.~H., {Tueller}, J., {Markwardt}, C.~B., {et~al.} 2013, ApjS,
  207, 19

\bibitem[{{Braito} {et~al.}(2011){Braito}, {Reeves}, {Sambruna}, \&
  {Gofford}}]{braito-2011}
{Braito}, V., {Reeves}, J.~N., {Sambruna}, R.~M., \& {Gofford}, J. 2011,
  \mnras, 414, 2739

\bibitem[{{Bruzual} \& {Charlot}(2003)}]{bruzual-charlot-2003}
{Bruzual}, G., \& {Charlot}, S. 2003, \mnras, 344, 1000

\bibitem[{{Davies} {et~al.}(2015){Davies}, {Burtscher}, {Rosario},
  {Storchi-Bergmann}, {Contursi}, {Genzel}, {Graci{\'a}-Carpio}, {Hicks},
  {Janssen}, {Koss}, {Lin}, {Lutz}, {Maciejewski}, {M{\"u}ller-S{\'a}nchez},
  {Orban de Xivry}, {Ricci}, {Riffel}, {Riffel}, {Schartmann},
  {Schnorr-M{\"u}ller}, {Sternberg}, {Sturm}, {Tacconi}, \&
  {Veilleux}}]{davies-2015}
{Davies}, R.~I., {Burtscher}, L., {Rosario}, D., {et~al.} 2015, \apj, 806, 127

\bibitem[{{Evans} {et~al.}(1991){Evans}, {Ford}, {Kinney}, {Antonucci},
  {Armus}, \& {Caganoff}}]{evans-1991}
{Evans}, I.~N., {Ford}, H.~C., {Kinney}, A.~L., {et~al.} 1991, \apjl, 369, L27

\bibitem[{{Evans} {et~al.}(2009){Evans}, {Beardmore}, {Page}, {Osborne},
  {O'Brien}, {Willingale}, {Starling}, {Burrows}, {Godet}, {Vetere}, {Racusin},
  {Goad}, {Wiersema}, {Angelini}, {Capalbi}, {Chincarini}, {Gehrels}, {Kennea},
  {Margutti}, {Morris}, {Mountford}, {Pagani}, {Perri}, {Romano}, \&
  {Tanvir}}]{evans-2009}
{Evans}, P.~A., {Beardmore}, A.~P., {Page}, K.~L., {et~al.} 2009, \mnras, 397,
  1177

\bibitem[{{Fabian} {et~al.}(2017){Fabian}, {Lohfink}, {Belmont}, {Malzac}, \&
  {Coppi}}]{fabian-2017}
{Fabian}, A.~C., {Lohfink}, A., {Belmont}, R., {Malzac}, J., \& {Coppi}, P.
  2017, MNRAS, 467, 2566

\bibitem[{{Fiore} {et~al.}(2009){Fiore}, {Puccetti}, {Brusa}, {Salvato},
  {Zamorani}, {Aldcroft}, {Aussel}, {Brunner}, {Capak}, {Cappelluti}, {Civano},
  {Comastri}, {Elvis}, {Feruglio}, {Finoguenov}, {Fruscione}, {Gilli},
  {Hasinger}, {Koekemoer}, {Kartaltepe}, {Ilbert}, {Impey}, {Le Floc'h},
  {Lilly}, {Mainieri}, {Martinez-Sansigre}, {McCracken}, {Menci}, {Merloni},
  {Miyaji}, {Sanders}, {Sargent}, {Schinnerer}, {Scoville}, {Silverman},
  {Smolcic}, {Steffen}, {Santini}, {Taniguchi}, {Thompson}, {Trump}, {Vignali},
  {Urry}, \& {Yan}}]{fiore-2009}
{Fiore}, F., {Puccetti}, S., {Brusa}, M., {et~al.} 2009, \apj, 693, 447

\bibitem[{{Gandhi} {et~al.}(2015){Gandhi}, {H{\"o}nig}, \&
  {Kishimoto}}]{gandhi-2015}
{Gandhi}, P., {H{\"o}nig}, S.~F., \& {Kishimoto}, M. 2015, \apj, 812, 113

\bibitem[{{Gandhi} {et~al.}(2009){Gandhi}, {Horst}, {Smette}, {H{\"o}nig},
  {Comastri}, {Gilli}, {Vignali}, \& {Duschl}}]{gandhi-2009}
{Gandhi}, P., {Horst}, H., {Smette}, A., {et~al.} 2009, \aap, 502, 457

\bibitem[{{Gehrels} {et~al.}(2004){Gehrels}, {Chincarini}, {Giommi}, {Mason},
  {Nousek}, {Wells}, {White}, {Barthelmy}, {Burrows}, {Cominsky}, {Hurley},
  {Marshall}, {M{\'e}sz{\'a}ros}, {Roming}, {Angelini}, {Barbier}, {Belloni},
  {Campana}, {Caraveo}, {Chester}, {Citterio}, {Cline}, {Cropper}, {Cummings},
  {Dean}, {Feigelson}, {Fenimore}, {Frail}, {Fruchter}, {Garmire}, {Gendreau},
  {Ghisellini}, {Greiner}, {Hill}, {Hunsberger}, {Krimm}, {Kulkarni}, {Kumar},
  {Lebrun}, {Lloyd-Ronning}, {Markwardt}, {Mattson}, {Mushotzky}, {Norris},
  {Osborne}, {Paczynski}, {Palmer}, {Park}, {Parsons}, {Paul}, {Rees},
  {Reynolds}, {Rhoads}, {Sasseen}, {Schaefer}, {Short}, {Smale}, {Smith},
  {Stella}, {Tagliaferri}, {Takahashi}, {Tashiro}, {Townsley}, {Tueller},
  {Turner}, {Vietri}, {Voges}, {Ward}, {Willingale}, {Zerbi}, \&
  {Zhang}}]{swift-mission}
{Gehrels}, N., {Chincarini}, G., {Giommi}, P., {et~al.} 2004, ApJ, 611, 1005

\bibitem[{{Ghisellini} {et~al.}(1993){Ghisellini}, {Haardt}, \&
  {Fabian}}]{corona-heating-paper}
{Ghisellini}, G., {Haardt}, F., \& {Fabian}, A.~C. 1993, \mnras, 263

\bibitem[{{Greene} \& {Ho}(2007)}]{green-ho-2007}
{Greene}, J.~E., \& {Ho}, L.~C. 2007, \apj, 670, 92

\bibitem[{{Harrison} {et~al.}(2013){Harrison}, {Craig}, {Christensen},
  {Hailey}, {Zhang}, {Boggs}, {Stern}, {Cook}, {Forster}, {Giommi},
  {Grefenstette}, {Kim}, {Kitaguchi}, {Koglin}, {Madsen}, {Mao}, {Miyasaka},
  {Mori}, {Perri}, {Pivovaroff}, {Puccetti}, {Rana}, {Westergaard}, {Willis},
  {Zoglauer}, {An}, {Bachetti}, {Barri{\`e}re}, {Bellm}, {Bhalerao},
  {Brejnholt}, {Fuerst}, {Liebe}, {Markwardt}, {Nynka}, {Vogel}, {Walton},
  {Wik}, {Alexander}, {Cominsky}, {Hornschemeier}, {Hornstrup}, {Kaspi},
  {Madejski}, {Matt}, {Molendi}, {Smith}, {Tomsick}, {Ajello}, {Ballantyne},
  {Balokovi{\'c}}, {Barret}, {Bauer}, {Blandford}, {Brandt}, {Brenneman},
  {Chiang}, {Chakrabarty}, {Chenevez}, {Comastri}, {Dufour}, {Elvis}, {Fabian},
  {Farrah}, {Fryer}, {Gotthelf}, {Grindlay}, {Helfand}, {Krivonos}, {Meier},
  {Miller}, {Natalucci}, {Ogle}, {Ofek}, {Ptak}, {Reynolds}, {Rigby},
  {Tagliaferri}, {Thorsett}, {Treister}, \& {Urry}}]{nustar-harrison}
{Harrison}, F.~A., {Craig}, W.~W., {Christensen}, F.~E., {et~al.} 2013, ApJ,
  770, 103

\bibitem[{{Hogg} {et~al.}(2012){Hogg}, {Winter}, {Mushotzky}, {Reynolds}, \&
  {Trippe}}]{hogg-2012}
{Hogg}, J.~D., {Winter}, L.~M., {Mushotzky}, R.~F., {Reynolds}, C.~S., \&
  {Trippe}, M. 2012, \apj, 752, 153

\bibitem[{{Kalberla} {et~al.}(2005){Kalberla}, {Burton}, {Hartmann}, {Arnal},
  {Bajaja}, {Morras}, \& {P{\"o}ppel}}]{galactic-nh}
{Kalberla}, P.~M.~W., {Burton}, W.~B., {Hartmann}, D., {et~al.} 2005, A\&A,
  440, 775

\bibitem[{{Kamraj} {et~al.}(2018){Kamraj}, {Harrison}, {Balokovi{\'c}},
  {Lohfink}, \& {Brightman}}]{nikita-2018}
{Kamraj}, N., {Harrison}, F.~A., {Balokovi{\'c}}, M., {Lohfink}, A., \&
  {Brightman}, M. 2018, \apj, 866, 124

\bibitem[{{Kara} {et~al.}(2017){Kara}, {Garc{\'{\i}}a}, {Lohfink}, {Fabian},
  {Reynolds}, {Tombesi}, \& {Wilkins}}]{kara-2017}
{Kara}, E., {Garc{\'{\i}}a}, J.~A., {Lohfink}, A., {et~al.} 2017, MNRAS, 468,
  3489

\bibitem[{{Koss} {et~al.}(2017){Koss}, {Trakhtenbrot}, {Ricci}, {Lamperti},
  {Oh}, {Berney}, {Schawinski}, {Balokovi{\'c}}, {Baronchelli}, {Crenshaw},
  {Fischer}, {Gehrels}, {Harrison}, {Hashimoto}, {Hogg}, {Ichikawa}, {Masetti},
  {Mushotzky}, {Sartori}, {Stern}, {Treister}, {Ueda}, {Veilleux}, \&
  {Winter}}]{koss-2017}
{Koss}, M., {Trakhtenbrot}, B., {Ricci}, C., {et~al.} 2017, \apj, 850, 74

\bibitem[{{Krolik} \& {Begelman}(1988)}]{krolik-1988}
{Krolik}, J.~H., \& {Begelman}, M.~C. 1988, \apj, 329, 702

\bibitem[{{Landi} {et~al.}(2007){Landi}, {Masetti}, {Morelli}, {Palazzi},
  {Bassani}, {Malizia}, {Bazzano}, {Bird}, {Dean}, {Galaz}, {Minniti}, \&
  {Ubertini}}]{landi-2007}
{Landi}, R., {Masetti}, N., {Morelli}, L., {et~al.} 2007, \apj, 669, 109

\bibitem[{{L{\'o}pez-Gonzaga} {et~al.}(2016){L{\'o}pez-Gonzaga}, {Burtscher},
  {Tristram}, {Meisenheimer}, \& {Schartmann}}]{ir-interferometry-2016}
{L{\'o}pez-Gonzaga}, N., {Burtscher}, L., {Tristram}, K.~R.~W., {Meisenheimer},
  K., \& {Schartmann}, M. 2016, \aap, 591, A47

\bibitem[{{Lusso} \& {Risaliti}(2016)}]{lusso-2016}
{Lusso}, E., \& {Risaliti}, G. 2016, \apj, 819, 154

\bibitem[{{Lusso} {et~al.}(2010){Lusso}, {Comastri}, {Vignali}, {Zamorani},
  {Brusa}, {Gilli}, {Iwasawa}, {Salvato}, {Civano}, {Elvis}, {Merloni},
  {Bongiorno}, {Trump}, {Koekemoer}, {Schinnerer}, {Le Floc'h}, {Cappelluti},
  {Jahnke}, {Sargent}, {Silverman}, {Mainieri}, {Fiore}, {Bolzonella}, {Le
  F{\`e}vre}, {Garilli}, {Iovino}, {Kneib}, {Lamareille}, {Lilly}, {Mignoli},
  {Scodeggio}, \& {Vergani}}]{lusso-2010}
{Lusso}, E., {Comastri}, A., {Vignali}, C., {et~al.} 2010, \aap, 512, A34

\bibitem[{{Lutz} {et~al.}(2004){Lutz}, {Maiolino}, {Spoon}, \&
  {Moorwood}}]{lutz-2004}
{Lutz}, D., {Maiolino}, R., {Spoon}, H.~W.~W., \& {Moorwood}, A.~F.~M. 2004,
  \aap, 418, 465

\bibitem[{{Madsen} {et~al.}(2018){Madsen}, {Harrison}, {Broadway},
  {Christensen}, {Descalle}, {Ferreira}, {Grefenstette}, {Gurgew},
  {Hornschemeier}, {Miyasaka}, {Okajima}, {Pike}, {Pivovaroff}, {Saha},
  {Stern}, {Vogel}, {Windt}, \& {Zhang}}]{madsen-2018}
{Madsen}, K.~K., {Harrison}, F., {Broadway}, D., {et~al.} 2018, in Society of
  Photo-Optical Instrumentation Engineers (SPIE) Conference Series, Vol. 10699,
  106996M

\bibitem[{Magdziarz \& Zdziarski(1995)}]{pexrav}
Magdziarz, P., \& Zdziarski, A.~A. 1995, MNRAS, 273, 837

\bibitem[{{Maiolino} {et~al.}(2001){Maiolino}, {Marconi}, {Salvati},
  {Risaliti}, {Severgnini}, {Oliva}, {La Franca}, \& {Vanzi}}]{maiolino-2001}
{Maiolino}, R., {Marconi}, A., {Salvati}, M., {et~al.} 2001, \aap, 365, 28

\bibitem[{{Marinucci} {et~al.}(2016){Marinucci}, {Bianchi}, {Matt},
  {Alexander}, {Balokovi{\'c}}, {Bauer}, {Brandt}, {Gand hi}, {Guainazzi},
  {Harrison}, {Iwasawa}, {Koss}, {Madsen}, {Nicastro}, {Puccetti}, {Ricci},
  {Stern}, \& {Walton}}]{marinucci-2016}
{Marinucci}, A., {Bianchi}, S., {Matt}, G., {et~al.} 2016, \mnras, 456, L94

\bibitem[{{Merloni} {et~al.}(2014){Merloni}, {Bongiorno}, {Brusa}, {Iwasawa},
  {Mainieri}, {Magnelli}, {Salvato}, {Berta}, {Cappelluti}, {Comastri},
  {Fiore}, {Gilli}, {Koekemoer}, {Le Floc'h}, {Lusso}, {Lutz}, {Miyaji},
  {Pozzi}, {Riguccini}, {Rosario}, {Silverman}, {Symeonidis}, {Treister},
  {Vignali}, \& {Zamorani}}]{merloni-2014}
{Merloni}, A., {Bongiorno}, A., {Brusa}, M., {et~al.} 2014, \mnras, 437, 3550

\bibitem[{Murphy \& Yaqoob(2009)}]{mytorus}
Murphy, K.~D., \& Yaqoob, T. 2009, MNRAS, 397, 1549

\bibitem[{{Oh} {et~al.}(2011){Oh}, {Sarzi}, {Schawinski}, \& {Yi}}]{oh-2011}
{Oh}, K., {Sarzi}, M., {Schawinski}, K., \& {Yi}, S.~K. 2011, \apjs, 195, 13

\bibitem[{{Oh} {et~al.}(2015){Oh}, {Yi}, {Schawinski}, {Koss}, {Trakhtenbrot},
  \& {Soto}}]{oh-2015}
{Oh}, K., {Yi}, S.~K., {Schawinski}, K., {et~al.} 2015, \apjs, 219, 1

\bibitem[{{Oh} {et~al.}(2018){Oh}, {Koss}, {Markwardt}, {Schawinski},
  {Baumgartner}, {Barthelmy}, {Cenko}, {Gehrels}, {Mushotzky}, {Petulante},
  {Ricci}, {Lien}, \& {Trakhtenbrot}}]{bat-105-month}
{Oh}, K., {Koss}, M., {Markwardt}, C.~B., {et~al.} 2018, \apjs, 235, 4

\bibitem[{{Panessa} \& {Bassani}(2002)}]{panessa-2002}
{Panessa}, F., \& {Bassani}, L. 2002, \aap, 394, 435

\bibitem[{{Panessa} {et~al.}(2006){Panessa}, {Bassani}, {Cappi}, {Dadina},
  {Barcons}, {Carrera}, {Ho}, \& {Iwasawa}}]{panessa-2006}
{Panessa}, F., {Bassani}, L., {Cappi}, M., {et~al.} 2006, \aap, 455, 173

\bibitem[{Perri {et~al.}(2017)Perri, Puccetti, Spagnuolo, Ficcadenti, Davis,
  Forster, Grefenstette, Harrison, \& Madsen}]{nustardas}
Perri, M., Puccetti, S., Spagnuolo, N., {et~al.} 2017, The NuSTAR Data Analysis
  Software Guide v1.7

\bibitem[{{Ricci} {et~al.}(2017){Ricci}, {Trakhtenbrot}, {Koss}, {Ueda}, {Del
  Vecchio}, {Treister}, {Schawinski}, {Paltani}, {Oh}, {Lamperti}, {Berney},
  {Gandhi}, {Ichikawa}, {Bauer}, {Ho}, {Asmus}, {Beckmann}, {Soldi},
  {Balokovi{\'c}}, {Gehrels}, \& {Markwardt}}]{ricci-2017a}
{Ricci}, C., {Trakhtenbrot}, B., {Koss}, M.~J., {et~al.} 2017, \apjs, 233, 17

\bibitem[{{Risaliti} {et~al.}(2005){Risaliti}, {Elvis}, {Fabbiano}, {Baldi}, \&
  {Zezas}}]{risaliti-2005}
{Risaliti}, G., {Elvis}, M., {Fabbiano}, G., {Baldi}, A., \& {Zezas}, A. 2005,
  \apjl, 623, L93

\bibitem[{{S{\'a}nchez} {et~al.}(2006){S{\'a}nchez}, {Garc{\'\i}a-Lorenzo},
  {Jahnke}, {Mediavilla}, {Gonz{\'a}lez-Serrano}, {Christensen}, \&
  {Wisotzki}}]{sanchez-2006}
{S{\'a}nchez}, S.~F., {Garc{\'\i}a-Lorenzo}, B., {Jahnke}, K., {et~al.} 2006,
  Astronomische Nachrichten, 327, 167

\bibitem[{{Shimizu} {et~al.}(2018){Shimizu}, {Davies}, {Koss}, {Ricci},
  {Lamperti}, {Oh}, {Schawinski}, {Trakhtenbrot}, {Burtscher}, {Genzel}, {Lin},
  {Lutz}, {Rosario}, {Sturm}, \& {Tacconi}}]{shimizu-2017}
{Shimizu}, T.~T., {Davies}, R.~I., {Koss}, M., {et~al.} 2018, \apj, 856, 154

\bibitem[{{Stern}(2015)}]{stern-2015}
{Stern}, D. 2015, \apj, 807, 129

\bibitem[{{Storchi-Bergmann} {et~al.}(1992){Storchi-Bergmann}, {Wilson}, \&
  {Baldwin}}]{storchi-1992}
{Storchi-Bergmann}, T., {Wilson}, A.~S., \& {Baldwin}, J.~A. 1992, \apj, 396,
  45

\bibitem[{{Str{\"u}der} {et~al.}(2001){Str{\"u}der}, {Briel}, {Dennerl},
  {Hartmann}, {Kendziorra}, {Meidinger}, {Pfeffermann}, {Reppin}, {Aschenbach},
  \& {Bornemann}}]{xmm-pn}
{Str{\"u}der}, L., {Briel}, U., {Dennerl}, K., {et~al.} 2001, \aap, 365, L18

\bibitem[{{Tortosa} {et~al.}(2017){Tortosa}, {Marinucci}, {Matt}, {Bianchi},
  {La Franca}, {Ballantyne}, {Boorman}, {Fabian}, {Farrah}, {Fuerst}, {Gandhi},
  {Harrison}, {Koss}, {Ricci}, {Stern}, {Ursini}, \& {Walton}}]{grs-low-ecut}
{Tortosa}, A., {Marinucci}, A., {Matt}, G., {et~al.} 2017, MNRAS, 466, 4193

\bibitem[{{Tran} {et~al.}(2011){Tran}, {Lyke}, \& {Mader}}]{tran-2011}
{Tran}, H.~D., {Lyke}, J.~E., \& {Mader}, J.~A. 2011, \apj, 726, L21

\bibitem[{{Trippe} {et~al.}(2011){Trippe}, {Reynolds}, {Koss}, {Mushotzky}, \&
  {Winter}}]{trippe-2011}
{Trippe}, M.~L., {Reynolds}, C.~S., {Koss}, M., {Mushotzky}, R.~F., \&
  {Winter}, L.~M. 2011, \apj, 736, 81

\bibitem[{{Tristram} {et~al.}(2007){Tristram}, {Meisenheimer}, {Jaffe},
  {Schartmann}, {Rix}, {Leinert}, {Morel}, {Wittkowski}, {R{\"o}ttgering},
  {Perrin}, {Lopez}, {Raban}, {Cotton}, {Graser}, {Paresce}, \&
  {Henning}}]{tristram-2007}
{Tristram}, K.~R.~W., {Meisenheimer}, K., {Jaffe}, W., {et~al.} 2007, \aap,
  474, 837

\bibitem[{{Turner} {et~al.}(2001){Turner}, {Abbey}, {Arnaud}, {Balasini},
  {Barbera}, {Belsole}, {Bennie}, {Bernard}, {Bignami}, \& {Boer}}]{xmm-mos}
{Turner}, M.~J.~L., {Abbey}, A., {Arnaud}, M., {et~al.} 2001, \aap, 365, L27

\bibitem[{{Urry} \& {Padovani}(1995)}]{urry-1995}
{Urry}, C.~M., \& {Padovani}, P. 1995, PASP, 107, 83

\bibitem[{Verner {et~al.}(1996)Verner, Ferland, Korista, \& Yakovlev}]{vern}
Verner, D.~A., Ferland, G.~J., Korista, K.~T., \& Yakovlev, D.~G. 1996, ApJ,
  465, 487

\bibitem[{{Walton} {et~al.}(2016){Walton}, {Tomsick}, {Madsen}, {Grinberg},
  {Barret}, {Boggs}, {Christensen}, {Clavel}, {Craig}, {Fabian}, {Fuerst},
  {Hailey}, {Harrison}, {Miller}, {Parker}, {Rahoui}, {Stern}, {Tao}, {Wilms},
  \& {Zhang}}]{walton-2016}
{Walton}, D.~J., {Tomsick}, J.~A., {Madsen}, K.~K., {et~al.} 2016, \apj, 826,
  87

\bibitem[{Wilms {et~al.}(2000)Wilms, Allen, \& McCray}]{wilm}
Wilms, J., Allen, A., \& McCray, R. 2000, ApJ, 542, 914

\bibitem[{{Winter} {et~al.}(2010){Winter}, {Lewis}, {Koss}, {Veilleux},
  {Keeney}, \& {Mushotzky}}]{winter-2010}
{Winter}, L.~M., {Lewis}, K.~T., {Koss}, M., {et~al.} 2010, \apj, 710, 503

\bibitem[{{Zdziarski} {et~al.}(1993){Zdziarski}, {Zycki}, \&
  {Krolik}}]{Zdziarski-1993}
{Zdziarski}, A.~A., {Zycki}, P.~T., \& {Krolik}, J.~H. 1993, \apjl, 414, L81

\end{thebibliography}

\end{document}